\documentclass[acmsmall]{acmart}

\AtBeginDocument{%
  \providecommand\BibTeX{{%
    \normalfont B\kern-0.5em{\scshape i\kern-0.25em b}\kern-0.8em\TeX}}}

\setcopyright{acmlicensed}
\acmJournal{PACMHCI}
\acmYear{2021} \acmVolume{5} \acmNumber{CSCW2} \acmArticle{335} \acmMonth{10} \acmPrice{15.00}\acmDOI{10.1145/3476076}

\acmConference[CSCW '21]{CSCW '21}{Virtual}{Online}
\acmBooktitle{CSCW '21: The 24th ACM Conference on Computer-Supported Cooperative Work and Social Computing, Virtual}
\acmPrice{15.00}
\acmISBN{978-1-4503-XXXX-X/18/06}
\usepackage{soul}
\usepackage{subcaption}
\usepackage{multirow}
\usepackage{csquotes}

\newcommand{\Comment}[1]{}

\newcommand{\chm}[1]{#1}

\usepackage{graphicx}
\usepackage{enumitem}
\setlist[description]{leftmargin=2\parindent,labelindent=2\parindent,rightmargin=2\parindent}

\setlist[itemize]{leftmargin=*}
\setlist[enumerate]{leftmargin=*}

\begin{document}

\title{Goldilocks: Consistent Crowdsourced Scalar Annotations with Relative Uncertainty}

\author{Quanze Chen}
\email{cqz@cs.washington.edu}
\affiliation{%
  \institution{University of Washington}
  \city{Seattle, WA}
  \country{USA}}

\author{Daniel S. Weld}
\email{weld@cs.washington.edu}
\affiliation{%
  \institution{University of Washington}
  \city{Seattle, WA}
  \country{USA}}
  
\author{Amy X. Zhang}
\email{axz@cs.washington.edu}
\affiliation{%
  \institution{University of Washington}
  \streetaddress{Sample Address}
  \city{Seattle, WA}
  \country{USA}}



\begin{abstract}
Human ratings have become a crucial resource for training and evaluating machine learning systems. However, traditional elicitation methods for absolute and comparative rating  suffer from issues with consistency and often do not distinguish between uncertainty due to disagreement between annotators and ambiguity inherent to the item being rated. In this work, we present Goldilocks, a novel crowd rating elicitation technique for collecting calibrated scalar annotations that also distinguishes inherent ambiguity from inter-annotator disagreement. We introduce two main ideas: grounding absolute rating scales with examples and using a two-step bounding process to establish a range for an item's placement. We test our designs in three domains: judging toxicity of online comments, estimating satiety of food depicted in images, and estimating age based on portraits. We show that (1) Goldilocks can improve consistency in domains where interpretation of the scale is not universal, and that (2) representing items with ranges lets us simultaneously capture different sources of uncertainty leading to better estimates of pairwise relationship distributions.
\end{abstract}

\maketitle

\section{Introduction}

Much of modern machine learning is built on the foundation of human-\chm{annotated} data. As \chm{the application of these models has expanded} into more socially embedded and \chm{contextually nuanced domains ~\cite{arora2020novel,mitra2015credbank,norregaard2019nela}, collecting high quality, consistent, and robust data from human annotators has become an increasingly important yet challenging task~\cite{bhuiyan2020investigating}. As one example, the ability to gather human evaluations of the toxicity of a piece of text is a necessary precursor to being able to build toxicity models to support online communities~\cite{wulczyn2017ex} as well as capture and mitigate harmful outputs generated by large language models~\cite{Gehman2020RealToxicityPromptsEN}.}

\chm{However, traditional rating methods commonly used today, like absolute or comparative rating, can produce inconsistencies in ratings across annotators and even with a single annotator's ratings~\cite{aroyo2019crowdsourcing,salminen2018online}. This is due to issues such as lack of a common interpretation of the scale in the case of absolute rating, as well as lack of global context in the case of comparative rating~\cite{Weijters2016TheCS,welty2019metrology,Clark2018WhyRW}.
Additionally, while current rating methods can capture uncertainty in the ratings, it is difficult to dissect whether the uncertainty is a result of inherent ambiguity in the item---where certain items cannot be confidently distinguished from each other~\cite{Dumitrache2018CapturingAI}---or from disagreement between annotators on where the item should be placed. 
Distinguishing these sources of uncertainty offers the potential of better capturing biases between annotators. It also allows us to develop more calibrated models that only make high-confidence distinctions between items when a human would have as well~\cite{Guo2017OnCO}.}

\begin{figure}
\begin{minipage}{\textwidth}
 \centering
 \includegraphics[width=0.95\linewidth]{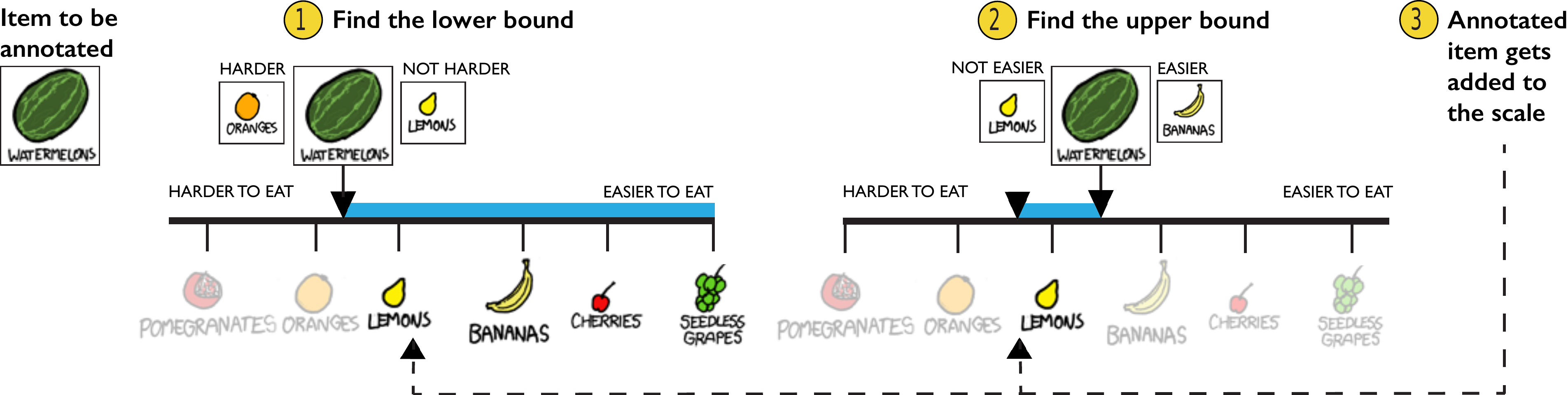}
 \captionof{figure}{The Goldilocks annotation process involves placing items onto a continuous scale that is populated with items that have previously been annotated. The process is broken down into three parts. (1) Find the lower bound by moving the left handle of the slider towards the right and away from its initial position on the far left of the scale. Continue until encountering an item on the scale that is either greater than or indistinguishable from the item to be annotated. (2) Find the upper bound in the same way but moving the right handle towards the left. Continue until encountering an item on the scale that is less than or indistinguishable from the item to be annotated or until the two handles are on top of each other, representing complete certainty. (3) Finally, the lower and upper bounds of the item get added to the scale to join the existing items. Thus, an annotator will be able to see and compare against their own prior annotated items as they annotate more items. Images of fruit are taken from XKCD: \url{https://xkcd.com/388/}}
 \label{fig:scale}
\end{minipage}
\end{figure}

In this paper, we propose a new design for collecting scalar annotations called Goldilocks\footnote{Somewhat like Goldilocks in ``Goldilocks and the Three Bears'', annotators must make use of \textit{multiple} comparisons.} that combines the ability to make direct comparisons between items with \chm{the simplicity of} a continuous absolute rating scale (Figure~\ref{fig:scale}). To accomplish this, Goldilocks \chm{introduces} two main ideas---(1) \textit{Calibration using Prior Annotations}: we provide previously annotated items as anchors to ground  interpretations of the scale both within and across annotators. (2) \textit{Item-level Resolution Elicitation using Ranges}: we use a two-step process to collect lower and upper bounds for each item instead of a single placement. Goldilocks combines strengths from both absolute and comparative ratings as annotators make multiple comparative judgments while placing an item on an absolute scale. In addition, by directly eliciting an annotator's own judgment of an item's inherent ambiguity instead of relying on aggregating inter-annotator agreement, Goldilocks can separate agreement from perceived ambiguity. 

To understand the effectiveness of these \chm{designs, we conducted three studies comparing aspects of the Goldilocks annotation process against traditional methods. In the first experiment, we evaluated \chm{whether anchoring scales with a shared set of previously annotated items can improve consistency of item placement across annotators. In the second experiment, we examined whether including an annotator's own prior annotations as anchors improves self-consistency. Our final experiment evaluated how well ranges captured using Goldilocks can recover the distribution of pairwise relationships as measured by traditional absolute and comparative rating.
Each of our experiments were conducted in three domains representative of the subjective or ambiguous rating tasks that can be challenging for traditional methods}: judging \textsc{toxicity} of online comments (short text), estimating \textsc{satiety} of food depicted in images (visual), and estimating \textsc{age} from portrait photos (visual). 
}

\chm{From the experiments examining anchors, we found that the addition of shared example anchors to ground rating scales improves rating consistency between annotators in domains where shared understanding of the scale is low.
We also found indications that showing one's prior annotations in a session as additional anchors may improve self-consistency on examples where there is high initial uncertainty. From the experiment examining ranges, we found that our two-step range annotation process allows us to infer pairwise relationship distributions that are more robust---simultaneously reflecting both uncertainty of single annotators and disagreement between annotators---compared to alternatives with a single value. Finally, we found that the size of range annotations provides an interpretation of uncertainty that is distinct from the uncertainty modeled via inter-annotator disagreement.}

We conclude with a discussion of \chm{the limitations and opportunities for Goldilocks.
Regarding efficiency, while our approach is more costly than performing just one of absolute or comparative rating, our method is cheaper than performing both, which would be necessary to recover the richer data that Goldilocks generates.
We discuss cases where a deeper understanding of uncertainty can be important for generating more trustworthy model predictions.
We also discuss what we envision as a scaled-up Goldilocks workflow: utilizing iterative improvement through multiple annotation sessions with designs for bootstrapping the initial set of anchors along with interesting problems to be explored in each of these aspects.
}

\section{Related Work}
In this section we review prior work on: \chm{(1) growing demand for consistent and robust human rating, (2) prior work building on absolute and comparative rating designs, (3) uncertainty and disagreement in crowd annotation, and (4) making use of uncertainty from human annotators in downstream machine learning tasks.}

\subsection{Demand for Improving Human Rating}

There is a growing demand for human annotation in domains involving ambiguous or subjective examples, largely due to rapid progress in machine learning. Human rating annotation has been used to create or validate a variety of training data, for example, in the domains investigating toxicity~\cite{wulczyn2017ex}, misinformation and credibility~\cite{bhuiyan2020investigating,mitra2015credbank}, and emotionally manipulative text~\cite{Huffaker2020CrowdsourcedDO}. However, there is also increasing concern for the robustness of datasets collected~\cite{welty2019metrology} and whether nuances like uncertainty are being represented~\cite{aroyo2015truth}.

Direct human rating of model output has also become prevelant in the evaluation of high performance models where automated metrics (e.g., BLEU, METEOR) start to fail~\cite{CallisonBurch2006ReevaluationTR,Agarwal2008MeteorMA, Denkowski2010ChoosingTR}. For example, human rating has been used to evaluate aspects of generative tasks (e.g., summarization, translation) in natural language processing by capturing characteristics like fluency, relevance, and conciseness which cannot be easily and reliably assessed with automated metrics~\cite{graham-etal-2013-continuous}. Human rating has also commonly been used to evaluate the output of chatbots~\cite{Sedoc2019ChatEvalAT} or to judge search results~\cite{google} or cluster quality~\cite{zhang2018evaluation}. Increasingly, human ratings (both comparative and absolute) are becoming an integral aspect in facilitating comparisons between models through evaluation leaderboards and shared tasks~\cite{specia-etal-2020-findings,khashabi2021genie}, where consistency and robustness of comparative results are crucial.

\subsection{\chm{Absolute and Comparative Rating Designs}}

\chm{One of the most common designs for collecting human ratings today is through \textit{absolute rating} scales, often in the form of Likert or semantic differential scales~\cite{likert1932technique,osgood1957measurement}. When a consistent interpretation of the scale can be established across annotators, designs based on absolute rating can offer many benefits such as being very efficient (only requiring a single annotation per item) and providing easily interpretable ratings that are globally contextualized (rather than depending on other items). However, many annotation domains do not have commonly accepted scales, meaning that divergent interpretations of a scale based on abstract text descriptions can become a source of disagreement and inconsistency across annotators~\cite{Weijters2016TheCS}. Even within an annotator's own annotations, the lack of a well defined scale means that to maintain consistent ratings, they must refer to their own memory of their past decisions which can be unreliable~\cite{Brown2007ATR}. Accounting for these inconsistencies requires additional effort---either through additional calibration~\cite{Gardner2007AnalyzingOS} or just identifying and reporting them~\cite{Geva2019AreWM}. Absolute scales can also be locally unreliable~\cite{welty2019metrology}---because items are only ever compared against the scale's anchors, pairwise comparisons between two items with similar values can only be rigorously done if the measurement resolution (uncertainty around the values) is also accounted for. }

\chm{As many consistency problems in absolute rating systems result from the lack of direct comparisons between actual items, a natural solution is to look towards the other major alternative---\textit{comparative ratings}~\cite{thurstone1927law}. In comparative rating systems, items are compared against against one another directly, circumventing the need for a scale as a proxy and providing highly reliable measurements of local relationships. This kind of comparison can also be more intuitive for annotators leading to comparative systems sometimes suggested as a more accurate alternative for ranking items~\cite{kiritchenko-mohammad-2016-capturing,Liang2020BeyondUS}. However, collecting comparative ratings can be considerably more costly (on the order of $N$ comparisons per item) unless sampling and ranking aggregation methods or partial comparisons, which trade off additional uncertainty, are used~\cite{Jin2020RankAV,kiritchenko-mohammad-2016-capturing}. The focus on local comparisons makes it easy for an annotator to inadvertently produce annotations that are not globally self-consistent, requiring post-hoc corrective action that may not reflect an annotator's actual judgment. Abandoning global context also means that if a rating score (rather than ranking) is desired, a numeric mapping like Elo rating needs to be done~\cite{Clark2018WhyRW}, which often come with assumptions about uniform spacing between items.}

\chm{Past work has explored hybrid approaches that combine aspects of comparative and absolute annotation. For example, Sakaguchi et al.~\cite{Sakaguchi2018EfficientOS} present EASL, a hybrid approach where items are rated using continuous absolute scales but similar items are grouped together for annotation allowing for some degree of comparison and contextualization. While similar in motivation, our work differs in that we make comparison an integral part of the annotation process rather than an optional source of context, allowing us to provide more consistency by grounding comparison against global anchors and capture uncertainty intuitively by using comparisons to establish bounds.}

\chm{Beyond the individual drawbacks mentioned above, neither of the two traditional annotation methods supports effective separation of the sources of uncertainty as a part of the the annotation process~\cite{Hullermeier2019AleatoricAE}. These sources include both aleatoric uncertainty, or irreducible ambiguity inherent to the item being rated, and epistemic uncertainty, or disagreement on the placement of the item. Absolute rating forces annotators to resolve inherent ambiguity into a precise placement causing both sources of uncertainty to be mixed. Meanwhile, comparative rating only provides an indirect view into inherent ambiguity through the size of equivalence sets. Separating the two sources of uncertainty is especially desirable as it can be an important tool for understanding properties of the items being annotated separate from biases or divergent interpretations among annotators.}

\subsection{\chm{Addressing Uncertainty and Disagreement}}

\chm{Uncertainty and disagreement has been a long recognized challenge when collecting crowdsourced annotations of all kinds. Early work in crowdsourcing focused on measuring perception-based objective aspects of items, taking the view that the uncertainty observed as disagreement between workers is the result of random noise from unreliable perception. To address this kind of uncertainty, methods such as majority voting, expectation maximization~\cite{Dawid1979MaximumLE}, Max-Margin Majority Voting~\cite{8423686} and even active learning based approaches~\cite{Lin2016Reactive} have been proposed, which attempt to improve the quality of the true measurement signal by aggregating across more annotators and accounting for the varying degree of noise introduced by each annotator. More recent lines of work recognize the deficiencies in single value answers, proposing instead to use answer distributions in the form of allowing multiple labels~\cite{Jurgens2013EmbracingAA, Dumitrache2015CrowdsourcingDF, Dumitrache2018CapturingAI} to capture the sources of uncertainty more comprehensively rather than attempt to remove it. Generally, these aggregation methods rely on large amounts of redundancy; thus, a major focus of prior work in this area has been in improving the efficiency of annotation work though collecting more information for each item or information about more items in each annotation task~\cite{chung2019efficient, kiritchenko-mohammad-2016-capturing}.}

Another view focuses on the idea that disagreements can arise from divergent interpretations of the data and task specification among workers~\cite{kairam2016parting,gordon2021de,aroyo2015truth} or even within an annotator as they are exposed to more data. One prior line of work, structured labeling~\cite{kulesza2014structured,chang2017revolt}, proposes that tools and techniques can be designed to assist people in reconciling the evolving interpretations of data both individually and collectively \chm{when labeling or generating taxonomies. Kuleza et al.~\cite{kulesza2014structured} note that maintaining consistency in annotation can be challenging even for experts. This motivated our exploration of improving self-consistency by incorporating past annotations.} 

Rubrics~\cite{yuan2016rubrics} and training (such as via gated instructions~\cite{liu-naacl2016}) have also been proposed as an effective way to unify understanding \chm{of the task requirements across workers. In fact, in practice, task designers oftentimes expend significant effort building detailed rubrics with complex training and gating processes.} Prior work \chm{in this area} has explored how to reduce such burdens on task designers through collaboratively creating rubrics with workers~\cite{Bragg2018SproutCT}. \chm{However, strict rubrics are often undesirable when the goal is to elicit human judgments on properties that are difficult to rigorously define such as those involving subjective interpretation.}

\chm{A softer form of rubrics can be made by using in-domain ground truth (or ``gold'') examples to anchor the interpretation of tasks including those involving scales. Gold solutions created by experts are often used during training~\cite{Doroudi2016TowardAL} in lieu of or in addition to instructions and rubrics. Reference examples can also be provided during the task such as in the MUlti Stimulus test with Hidden Reference and Anchor (MUSHRA test)~\cite{Vlker2018ModificationsOT}. However, existing methods depend on curated or synthesized fixed gold anchors ahead of labeling, which, in the case of scale anchors for subjective domains, still requires a concrete definition of the scale ahead of time. Fixed anchors are also limited in the support they can provide for self-consistency.}

Finally, on challenging or high-stakes domains where correctness is important, deliberation has been proposed as a way of addressing and resolving disagreement directly. Deliberation processes can range from simple one-shot reconsideration prompts~\cite{drapeau2016microtalk} to more complex multi-turn discussions~\cite{chen2019cicero, schaekermann2018resolvable}. Alternatively, lighter methods have been proposed that model behavior of humans to identify when disagreement is likely~\cite{gurari2017crowdverge}. \chm{However, these methods can still require significant worker effort. }

\subsection{Recognizing and Leveraging Uncertainty in ML}

As it is impossible to eliminate uncertainty~\cite{Hullermeier2019AleatoricAE}, downstream tasks like machine learning have started to explore paths of utilizing uncertainty information (when it is available) during training and evaluation. Many machine learning models have been built to do tasks like classification~\cite{Bouveyron2009RobustSC}, ranking~\cite{Yan2007RankingWU} or regression~\cite{Yan2008RegressionFU} using uncertain labels.

Recently, there is growing interest in understanding and mitigating adversarial attacks on machine learning~\cite{Goodfellow2015ExplainingAH}. These attacks often trick models to make high confidence predictions that are incorrect due to limited ability for many models to accurately model its own confidence. One potential mitigation strategy has been to look at improving a model's robustness by improving ability to model uncertainty~\cite{Qin2020ImprovingUE}. Additionally, there is a push for models to more accurately understand when it should be unsure~\cite{Rajpurkar2018KnowWY}. These all motivate an increasing need to understand what humans find uncertain and calibrated ways for humans to convey their degree of uncertainty.

\section{Design}

Absolute rating can suffer from inconsistent scale interpretations while comparative rating lacks global context. Our design for the Goldilocks annotation system takes a hybrid approach, with the specific goals of: (1) improving consistency (between annotators and over time within annotator), and (2) enabling intuitive indication of uncertainty with respect to the scale for each example being labeled. 

In this section, we will describe the designs that address each of the goals above followed by additional aspects of operating the complete annotation workflow. At the end, we will \chm{discuss specific details of the design decisions we made for our implementation} separate from the overall design of the Goldilocks annotation process. 

\begin{figure}
\begin{minipage}{\textwidth}
 \centering
 \includegraphics[width=.90\linewidth]{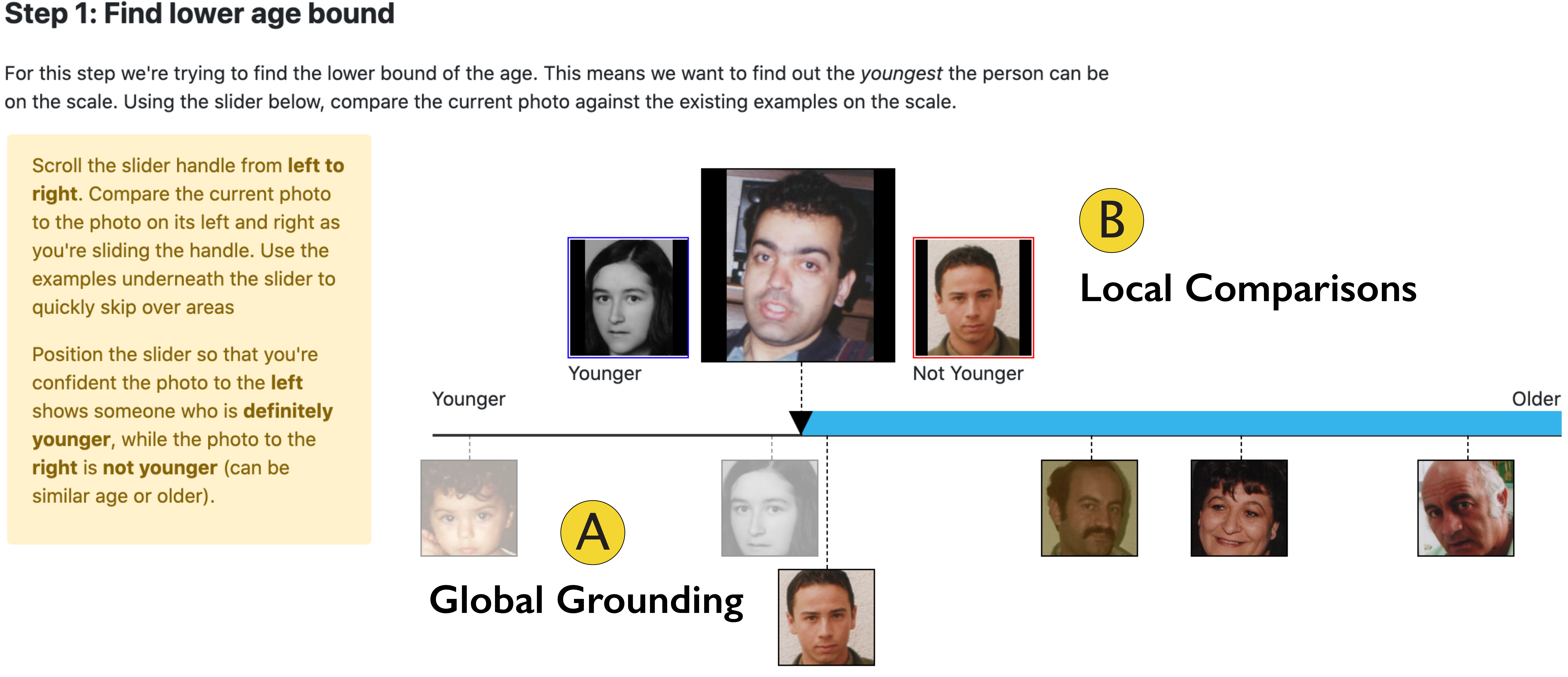}
 \captionof{figure}{A screenshot showing the comparisons that annotators can make while placing the upper or lower bound of an item on the scale in the Goldilocks annotation process. To support grounding with examples, Goldilocks provides: (A) global grounding by selecting 5--7 previously annotated items that are maximally spread out on the scale and placing them as anchors to support coarse and fast global adjustment. (B) Local comparisons of previously annotated items directly to the left and right of the slider handle are shown as an annotator scrubs the handle across the slider. Local items that are not one of the global examples are inserted as anchors. Together, this allows annotators to make fine-grained local adjustments.}
 \label{fig:comparisons}
\end{minipage}
\end{figure}

\subsection{Grounding with Prior Examples}

We base the main interactions in Goldilocks around an absolute rating design. To mitigate the aforementioned drawbacks of absolute rating, Goldilocks uses prior examples in addition to abstract descriptions to ground the scale, making it possible to make pairwise comparisons while still using absolute rating interactions. Prior work has shown that human judgments measured explicitly with comparisons can be easier than direct labels for some tasks~\cite{Simpson2018FindingCA, Zhang2017QuantifyingFA, Wah2014SimilarityCF}, and \textit{fixed} reference anchors have been used in other procedures to provide a more concrete grounding of scales~\cite{Vlker2018ModificationsOT}. Similar ideas that use comparisons against samples to contextualize abstract scales also exist in other fields like cognitive psychology~\cite{Stewart2006DecisionBS}.

Goldilocks uses a set of previously-annotated examples to add two additional pieces of information to the absolute rating scale---\textbf{global grounding} and \textbf{local comparisons}, as shown in Figure~\ref{fig:comparisons}. With \textbf{global grounding}, a small set of representative examples are selected and placed as anchors along the rating scale, similar to existing \chm{text-based} anchors for levels in \chm{traditional} absolute rating. Using concrete examples allows annotators to quickly understand and estimate where each item could fit on the scale. \chm{Since there can be many previously-annotated examples,} we make sure to only \chm{visualize a smaller subset} of examples (around 5 to 7, similar to typical numbers of Likert levels) that are maximally spread out along the scale. \chm{In practice, there are many ways to select these examples. The specific selection process we used is outlined in \ref{sec:implementation}.} 

While global grounding is useful for making coarse placements, it alone is insufficient for narrowing down specific placement of items. To help the annotators find specific placements, Goldilocks also surfaces \textbf{local comparisons} \chm{by showing} the immediate neighborhood above and below a position on the scale. As annotators scrub along a continuous scale, we show side-by-side comparisons between the \chm{current indicated position and the closest items above and below this position. Placements of these neighbors are also indicated on the scale itself}, allowing for annotators to adjust proportional distance to each neighbor based on \chm{their evaluation of the item being placed.} These designs together allow for a more \chm{consistent} and concrete \chm{instantiation} of the scale across multiple annotators.

\chm{Finally, Goldilocks addresses local self-consistency by supporting dynamic augmentation of the anchor examples used to ground the rating scale: as annotators progress in an annotation session, their own annotations for earlier items are also incorporated into the set of references alongside any pre-seeded ones (Step 3 of Figure~\ref{fig:scale}). These personal annotations will then also take part in both global grounding and local comparisons, making it possible to directly compare new items against past annotations produced in the same session.}

\chm{One potential limitation for any annotation process involving examples is how to start the annotation when no past examples are available. Goldilocks accounts for this with a separate procedure to curate an initial seed set that is deployed when past examples do not exist. } We will dive into more detail about the selection of \chm{this} initial seed set of items to jumpstart annotation in Section \chm{\ref{sec:cold-start-procedure}}. \chm{In the discussion section, we will also discuss avenues of addressing other challenges in example-based grounding such as scaling up annotation with iterative improvement and addressing density as the scale becomes populated with more annotated examples.}

\subsection{Two-Step Range Annotation}
\label{sec:twostep}

Not all items can be meaningfully distinguished from all other items by an annotator. Instead of forcing the breaking of ties, most designs for side-by-side comparisons allow annotators to indicate ``indistinguishable'' or ``tied'' pairs~\cite{Lubli2018HasMT}---however, there is no such elicitation process for traditional absolute rating designs.
With Goldilocks, we propose a new process that allows annotators to indicate ``indistinguishable'' pairwise relationships on an absolute rating scale. To achieve this, we take inspiration from prior work~\cite{Dumitrache2018CrowdsourcingGT}, where annotators were asked to select \textit{all} potentially relevant labels for an item instead of a single best label option. We extend this into the continuous scale domain by introducing the concept of eliciting ``range'' labels---where upper and lower bounds establish a subsection of the scale representing where an item \textit{could} be placed. Our range-based approach is also reminiscent of methods like best-worst scaling~\cite{kiritchenko-mohammad-2016-capturing} in comparative rating, which can efficiently capture pairwise relationships across many items. 

Prior designs have explored \chm{alternatives to eliciting uncertainty for scalar annotations, such as in the form of a weighted distribution across \textit{surrounding} anchor labels~\cite{chung2019efficient}. However, estimating distributions in this way can be challenging for humans, as an annotator has little guidance on how to allocate weight to the anchoring labels they find reasonable.} In Goldilocks, we \chm{can take advantage of the comparisons affored by grounding examples to contextualize distributions intuitively. Specifically, we break down the process of eliciting ranges into two steps: finding the lower bound and then finding the upper bound (Steps 1 and 2 in Figure~\ref{fig:scale}).} In the first step, an annotator \chm{can utilize the past example anchors to quickly search for where to place the lower bound of an item using comparisons to work up the scale and finding the position where they can no longer confidently decide that the closest reference should be lower on the scale than the annotated item.} Similarly, in the second step, an annotator establishes the upper bound working down from the scale and stopping when they can no longer identify a reference item as higher than the annotated item. 

Positions of anchor items on the scale are themselves \chm{internally represented by ranges. During each step, the anchors are visualized using the corresponding opposing bound: when finding the \textit{lower} bound for an item, anchor items are placed on the scale according to their \textit{upper} bound values and vice versa for the upper bound (shown in Figure~\ref{fig:scale}). This two-step process allows an annotator to easily establish a range that is intuitive and meaningful---it represents the range where the annotator is no longer able to confidently distinguish items.}

\begin{figure}
\begin{minipage}{\textwidth}
 \centering
 \includegraphics[width=1.0\linewidth]{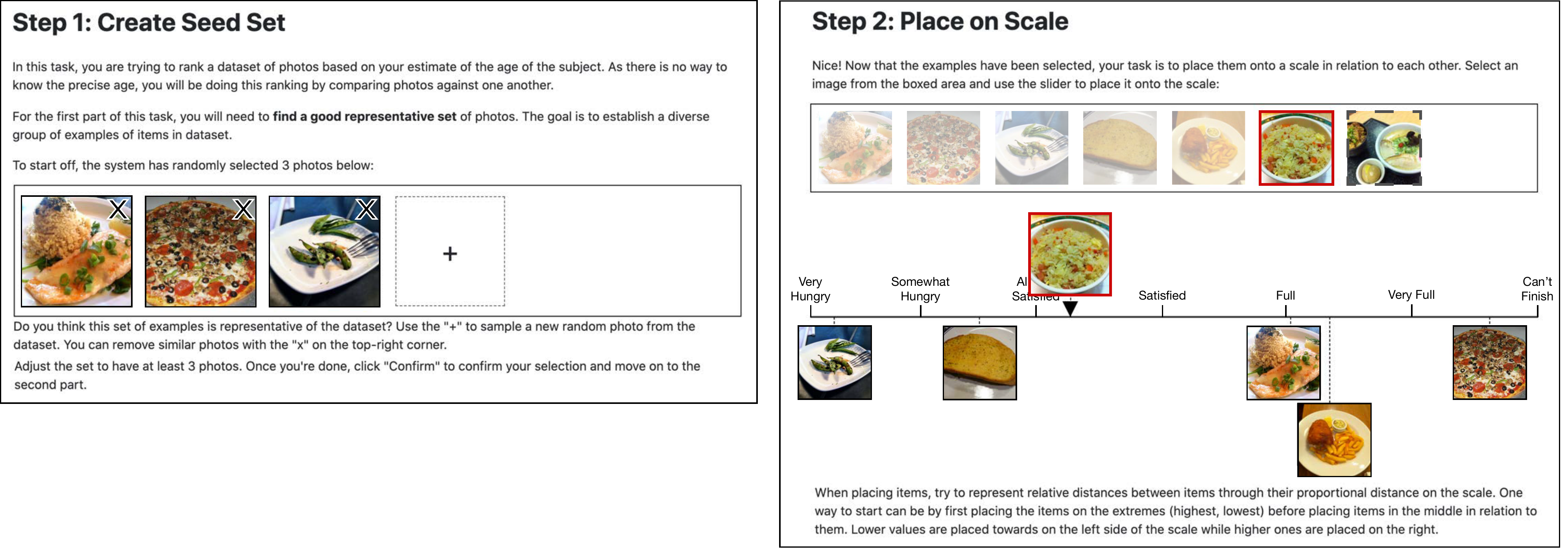}
 \captionof{figure}{Screenshots illustrating the two steps in the cold start process for Goldilocks. Step 1 (Left): A seed set can be created by using the cold start interface to randomly draw examples and drop existing ones to create an adequately sized representative set of examples. Step 2 (Right): The items from the seed set are placed onto a scale by adjusting their position relative to each other, forming the initial values that can be used to bootstrap the annotation tasks in Goldilocks. These initial items can later be reintroduced in the Goldilocks annotation process once other items have been annotated, in order to collect ranges.}
 \label{fig:cold_start_process}
\end{minipage}
\end{figure}

\subsection{Cold Start Process} \label{sec:cold-start-procedure}
Annotation of any item in the Goldilocks process requires there to be previously-annotated items using the same scale in order to populate the grounding examples and comparisons. 
However, if prior annotations do not exist yet, they must be created in what we call the cold start process. 

The cold start process (shown in Figure~\ref{fig:cold_start_process}) consists of two steps---representative example selection and placement on a scale. In the example selection step, Goldilocks draws a certain amount of un-annotated examples randomly from the set of data to be annotated. An annotator can then adjust this set by requesting to draw additional random examples or drop\chm{ping} existing examples. The goal is to adjust this set to be \chm{more} representative such that there are at least a certain number of examples in the set (defined based on task) and that the examples are maximally different from each other. \chm{A similar sample and replace approach was used in Alloy~\cite{chang2016alloy} to bootstrap good seed sets for clustering.} In the placement step, the annotator successively places all the examples onto an absolute rating scale by comparing them against each other, with the ability to adjust the position of any item on the scale. The scale can be blank at the outset or be initialized with text anchors as shown in Figure~\ref{fig:cold_start_process}. 

\chm{The cold start process can be completed with recruited annotators, where the resulting placements are aggregated across them to create the set of seed examples that become the first set of Goldilocks example anchors. Alternatively, the cold start process can be completed by the task designer or by domain experts, making it a way for requesters to specify a scale without having to design a set of training instances.} In this case, the steps \chm{in the cold start process} are used to assist the exploration of the dataset. Once additional items have been annotated using Goldilocks, the \chm{set of anchor examples can be augmented with this newly annotated data. If desired, the initial seed examples can be re-annotated by removing them from the scale and re-introducing them as new items to be placed in an iterative improvement fashion.}

\subsection{Implementation Details} \label{sec:implementation}
We outline specific details about our implementation of Goldilocks that we use for experiments. We implemented Goldilocks based on a custom slider component using JavaScript, HTML, and CSS. Global grounding examples were \chm{incorporated as part of the scale via fixed anchor tick markers below the scale}. Examples were then rendered in a fixed size box attached to each \chm{tick mark}. Images \chm{were scaled} to cover the box, and short text \chm{was presented as} as scrollable content within each box (Fig~\ref{fig:comparisons}). \chm{The interface selects global grounding examples by sorting the set of potential examples and progressively selecting examples that are at least a certain minimum distance from each other. As an annotators scrolls the slider handle}, we dynamically \chm{search for immediate} neighbor examples \chm{above and below the slider position} and render them as \chm{additional anchors placed} among the global grounding examples. \chm{Neighbor examples are also placed next to the item being annotated to facilitate comparison. Vertical positioning of the rendered anchor examples is dynamically adjusted so that they never visually overlap with each other.}

As our experiments were conducted on the Amazon Mechanical Turk crowdsourcing platform, we also implemented a gated training~\chm{\cite{liu-naacl2016}} phase for each of the annotation experiments. This phase \chm{focuses on} training the crowd workers to use the annotation interface rather than \chm{annotating a specific task domain, so we used a} common training example based on age estimation across all domains. Workers are presented with a prompt describing the task and interface, including specific actions \chm{that can be performed} using the interface. As workers complete \chm{each annotation} step \chm{for the training} task, we check their partial answers \chm{against the reference and provide just-in-time feedback} if they make a mistake. Once the worker accurately completes the training task, they will \chm{progress into the actual annotation task and given the specific instructions for the domain they are annotating}. We implemented \chm{some basic} quality control measures to prevent gaming of the task \chm{such as requiring workers to have interacted with the slider before they are allowed to proceed onto the next item.}

\section{Experiments}

In order to answer the research questions behind our Goldilocks designs, we conducted annotation experiments using data from 3 domains on the Amazon Mechanical Turk (AMT) platform and using interfaces that isolate specific aspects of Goldilocks for experimentation. Specifically, we tested the following hypotheses:

\begin{itemize}
    \item RQ1: Does grounding with examples improve consistency?
    \begin{itemize}
        \item H1-a: Using example-based anchors reduces the amount of disagreement between annotators on ratings of items compared with using semantic text descriptions as anchors.
        \item H1-b: Including an annotator's own annotations from the session as additional anchors results in improved self-consistency reflected by less disagreement with their past placement when placing items again.
    \end{itemize}
    \item RQ2: Does the range-based process create robust output for understanding relationships between items?
    \begin{itemize}
        \item H2-a: Range annotation captures item resolution and thus can more accurately model distributions of pairwise relationships (more than, less than, indistinguishable) compared to distributions produced by comparing single value annotation output.
        \item H2-b: Resolution of items captured using range annotation are better for modeling pairwise relationships than resolution captured through inter-annotator (dis)agreement.
    \end{itemize}
    \item RQ3: Does the uncertainty about items captured through the size of the ranges correlate with uncertainty captured in the form of inter-annotator disagreement in traditional semantic scale absolute ratings?
\end{itemize}

\subsection{\chm{Annotation Task Design}} \label{sec:studies}

\chm{
We describe in more detail the task design we used in our annotation experiments, including interfaces derived from Goldilocks and ones from traditional annotation. Unique crowd workers were recruited to use one of the following interfaces to provide annotations for a group of examples: 
}

\begin{figure}
\begin{minipage}{\textwidth}
 \centering
 \includegraphics[width=1.0\linewidth]{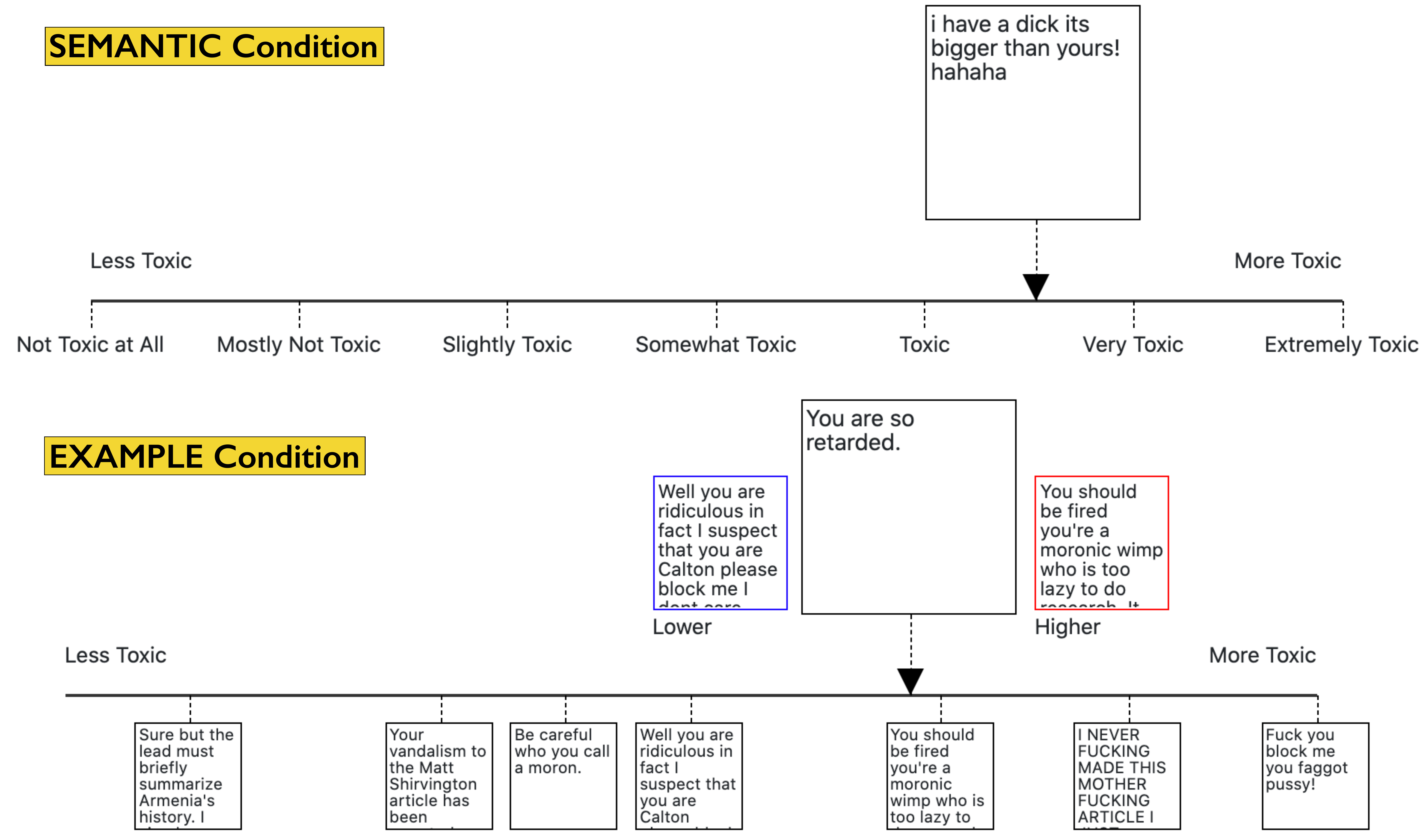}
 \captionof{figure}{Screenshot showing the two interfaces conditions (top: \textsc{semantic}, and bottom:\textsc{example}) used to evaluate consistency consistency between annotators. Examples shown in figure are from the \text{toxicity} domain pilot tasks.}
 \label{fig:between_worker_comparisons}
\end{minipage}
\end{figure}

\begin{itemize}
    \item \chm{\textbf{Single Value with Semantic Anchors (SV-SA)}: In each step, annotators are are asked to find a slider position that represents the placement of one item in the annotation sequence using a semantic scale as reference (Figure~\ref{fig:between_worker_comparisons} top).}
    \item \chm{\textbf{Single Value with Example Anchors (SV-EA)}: In each step, annotators are asked to find a slider position that represents the placement of one item in the annotation sequence using a scale anchored by other example item instances (Figure~\ref{fig:between_worker_comparisons} bottom). Depending on the experiment and condition, the annotator's past placements in earlier steps may become additional anchors for steps in the future.}
    \item \chm{\textbf{Pairwise}: Annotators were asked to compare all pairs of items. For each step in the annotation sequence, an annotator was presented with 1 reference item and a list of items it has not been compared to yet. For each item, the annotator was asked to judge the relationship of that item compared with the reference item ($>$,$<$,$\approx$). }
    \item \chm{\textbf{Range with Hybrid Anchors (R-HA)}: This represents the full proposed Goldilocks design. Annotators are given both semantic labels and example instances as reference anchors. For each item, an annotator is first asked to place a lower bound marker for the item followed by placing an upper bound marker. Ranges annotated in earlier steps are incorporated as additional anchors.}
\end{itemize}

\chm{Our first study (\ref{sec:consistency-between}) examines whether example anchors (\textbf{SV-EA}) improve agreement between annotators compated to semantic anchors (\textbf{SV-SA}). Following that, our second study (\ref{sec:consistency-within}) examines whether including an annotator's past placements improves within annotator consistency when using the \textbf{SV-EA} annotation design. Finally, in our last study (\ref{sec:ranges}), we collect ground truth pairwise relationships directly using the \textbf{Pairwise} interface, and compare how well we can recover the distribution of these relationships using data from the traditional single-value semantic anchor approache (with \textbf{SV-SA}) with that of the full Goldilocks range annotation design (\textbf{R-HA}).}

\chm{In all cases, annotators were first given a brief gated ``interface training'' instructional stage where they are guided to annotate a single item (based on an age estimation domain) using the annotation interface they were assigned. Instructions are provided during the process to guide them through using the interface and feedback is given if the annotator makes a mistake in the annotation. Once an annotator completes the annotation process without mistake, they are given details about the actual task domain they are annotating. Each annotator is then prompted to annotate a sequence of items using the assigned condition's interface.}

\subsection{Annotation Domains and Datasets}
We selected the following 3 annotation domains to to conduct annotation tasks: \textsc{toxicity}, \textsc{satiety} and \textsc{age}. These domains were selected to represent common types of rating tasks that have subjective aspects where a Goldilocks style approach to annotation could be desirable. These tasks also span two different modalities, short text and image, which closely align with rating tasks commonly conducted. 

\subsubsection{Toxicity} 

For this task domain, annotators judge the degree of toxicity in a short online comment, estimating how strongly the author of the comment intended to offend. Research has demonstrated that human judgments of online toxicity vary considerably from rater to rater due to subjectivity of the task~\cite{salminen2018online}. The \textsc{toxicity} domain represents a short text annotation task where annotators compare pieces of text that only consist of a couple of sentences. Similar tasks include judging fluency of text generation or judging text sentiment. To produce the annotation dataset for this domain, we sampled a 50:50 label-balanced subset of 100 comments from the Jigsaw comment toxicity classification challenge dataset~\cite{wulczyn2017ex} behind the Perspective API\footnote{\url{https://www.perspectiveapi.com}} which contains Wikipedia comments and binary labels of toxicity. Only comments that had between 4 and 280 characters (after markup removal) were sampled.  
When presenting the task to crowd workers, we borrow Perspective API's definition of a toxic comment: `a rude, disrespectful, or unreasonable comment that is likely to make you leave a discussion'. We also contrastively define healthy comments as those `relevant to the discussion' and further note that comments `can express disagreement'.

\subsubsection{Satiety} 

For this task domain, annotators judge how filling (satiable) is the food depicted in an image, taking into account the type of food and the portion size. The \textsc{satiety} domain represents an annotation task that contains uncertainty in the visual modality. Prior research has shown that while pairwise comparisons of food for expected satiety can result in robust ratings, personal familiarity also resulted in biases~\cite{brunstrom2008measuring}. We produced the annotation dataset by selecting a subset of food types from the Food-101 dataset~\cite{Bossard2014Food101M} and then sampling images for each selected food type up to a total of 80. One round of manual inspection was also done to verify food was clearly discernable in all images. 

\subsubsection{Age} 

For this task domain, annotators estimate the age of the subject depicted in a photo. The \textsc{age} domain is another annotation task in the visual modality that contains uncertainty, however age is grounded to a concrete scale that we expect most people to be already familiar with. We produced the annotation datset by sampling a subset of 100 portrait images from the FG-NET face dataset~\cite{ranking2014ECCV}.

\subsection{Anchors for each Domain} \label{sec:anchor-defn}

To maintain consistency across experiments, we defined a set of \chm{text-based semantic differential scale anchors} and a set of example anchors for each domain that was held constant across experiments. \chm{For the semantic scale anchors, we used text descriptions similar to 7-point Likert or semantic differential scales. Example anchors consisted of 7 roughly evenly spaced in-domain items each associated with a position on the scale.} 

For the \textsc{toxicity} domain, we \chm{used the following text descriptions for} semantic scale levels: ``1 - Not Toxic at All'', ``4 - Somewhat Toxic'' and ``7 - Extremely Toxic''. Other levels (2, 3, 5, 6) on the scale were presented as \chm{a} number without any associated description. The 7 example anchors were manually picked from a set of annotated examples produced from a pilot run of the cold start process with crowd annotators. 

For the \textsc{satiety} domain, we \chm{used the following text descriptions for} semantic scale levels: ``1 - Very Hungry', ``2 - Somewhat Hungry'', ``3 - Almost Satisfied'', ``4 - Satisfied', ``5 - Full'', ``6 - Very Full'', and ``7 - Can't Finish''. The 7 example anchors were produced by the authors producing gold annotations directly using the cold start process interface.

For the \textsc{age} domain, we \chm{used text scale levels based on numeric age values} ranging from ``0'' to ``60+'' incrementing in steps of 10.  The 7 example anchors were picked by finding all images corresponding to each semantic age level and then drawing a random one at each level and assigning its value to be the ground truth age. 

\subsection{\chm{Crowd Annotator Recruitment and Compensation}}

\chm{We recruited annotators for our experiments from the Amazon Mechanical Turk crowdsourcing platform from the United States with the qualification of approval rate no lower than 90\% and over 1000 approved HITs completed in the past. Across all studies, annotators were only allowed to participate in annotation if they had both not used the corresponding interface and not annotated the domain before. Overall, we recruited 655 unique workers across all 3 studies with an additional 44 unique workers who only participated in the pairwise annotation used to establish the ground truth for Study 3. For all annotation tasks, we set a base pay of \$0.10 which was given if the worker completed the training phase. Remaining compensation was distributed in the form of a bonus based on the interface being used and the number of items annotated.}

\chm{ Participants assigned to the \textbf{Single Value} tasks (both with \textbf{Semantic} and \textbf{Example} anchors) were given a per-item bonus of \$0.03 (for annotating a group of 10 or 20 items). Participants assigned to the \textbf{Range} tasks were given a per-item bonus of \$0.05 (a total of 10 items). Participants assigned to the \textbf{Pairwise} annotation tasks were given a per-comparison bonus of \$0.01 (a total of 45 comparisons). 
We set pay based on our estimate of time needed taken from pilot studies and used completion bonuses to correct for any discrepancies.
Based on condition, a final completion bonus of \$1.00, \$0.50, or \$1.00 for each of the previously mentioned interfaces respectively was provided. We distributed the final bonus in 2 batches as the initial completion bonus values we set for the tasks resulted in a measured hourly pay that was lower than desired. The final hourly rate measured between \$9.70 and \$10.90 across the various domains and interfaces when assuming the median work time for each interface. }

Manual quality checks were conducted on cases with a large number of similarly annotated values across different items (e.g., consistently placing at 0 or 1) as well as abnormally short work time, resulting in removal of 5 workers (and re-collection of corresponding annotations) across all experiments. \chm{Removed workers were included in the counts of recruited workers above. Within the removed workers, those} intentionally spamming across their entire sequence of annotations \chm{(choosing the exact same placement for all items)} only received the base pay for the task.

\subsection{Study 1: Evaluating Consistency Between Annotators} \label{sec:consistency-between}

We first explore whether example-based grounding presented in Goldilocks can improve consistency between different annotators (H1-a). \chm{For this experiment, we assigned each annotator to one of two conditions: \textsc{semantic}, where they were given 7-point text-based semantic anchors and presented with the \textbf{SV-SA} interface; or \textsc{example}, where they are given 7 example instances placed onto the scale using the \textbf{SV-EA} interface. For each domain, the anchors used are detailed in \ref{sec:anchor-defn}. We drew example anchor instances for the \textsc{toxicity} and \textsc{satiety} domains from past pilots of semantic differential scale annotation on a disjoint set of items, using average rating to establish their initial placement.  For the \textsc{age} domain, example instances were selected from a separate set of images drawn from the same dataset using the included ground truth age labels for initial placement.}

\chm{After the training, each annotator was tasked with} annotating a sequence of 10 items using the interface of the condition they were \chm{assigned}. To create sequences, each domain's dataset was shuffled once and then partitioned \chm{into equal-sized disjoint sets}. Each sequence for each domain was annotated by 10 workers in each of the two conditions. Annotators' placements of items on the scale was mapped as a continuous numeric value within the range [0, 1]. For the \textsc{toxicity} domain, the first and last items in each sequence were set to the same item to pilot measurement of within-annotator consistency, so only the 8 remaining annotations were used for analysis \chm{in this experiment}.

\subsubsection{Results}

\begin{table}[t]
    \centering
    \caption{Results for the experiment measuring consistency between annotators comparing between \textsc{semantic} and \textsc{example} conditions. Average disagreement is calculated as the standard error (over 10 annotators) for each instance averaged across all annotated instances. Significance testing done as a paired t-test across conditions for disagreement. \chm{We also examine how much of the 0-1 scale is being used by annotators on average in each condition by averaging each annotator's minimum and maximum rating values.}}
    \label{tab:results_consistency}
    \begin{tabular}{llccl}
    \toprule
    \textbf{Domain}  & \textbf{Condition}  & \textbf{Avg. Disagreement} & \textbf{Significance} & \textbf{\chm{Scale Util. ($\overline{\text{Min}}$, $\overline{\text{Max}}$)}} \\
    \midrule
    \textsc{toxicity} & \textsc{semantic} & 0.07348 & P < 0.001        &  \chm{0.773 (0.103, 0.876)}     \\
                      & \textsc{example}  & \textbf{0.06379} & Very Significant &  \chm{0.794 (0.104, 0.899)}        \\
    \midrule
    \textsc{satiety} & \textsc{semantic}  & 0.06373 & P < 0.005        &  \chm{0.603 (0.230, 0.833)}        \\
                     & \textsc{example}   & \textbf{0.05548} & Significant      &  \chm{0.635 (0.166, 0.801)}       \\
    \midrule
    \textsc{age}     & \textsc{semantic}  & \textbf{0.02765} & P < 0.001           &  \chm{0.696 (0.054, 0.751)}       \\
                     & \textsc{example}   & 0.04443 & Very Significant    &  \chm{0.593 (0.072, 0.665)}      \\
    \bottomrule
    \end{tabular}
\end{table}

To evaluate the amount of consistency between annotators for each annotated data point, we computed the standard error across annotators as a proxy for the amount of disagreement. We note that the standard error values are comparable across conditions as the range of values on the scale and number of annotators was fixed between all conditions. We also evaluated the significance of any difference by conducting a two-tailed paired t-test on the standard error of each annotated item across each pair of conditions (\textsc{semantic} versus \textsc{example}) in each domain. A summary of the results are shown in Table~\ref{tab:results_consistency}.

\begin{figure}[tbh]
    \centering
    \begin{subfigure}{.32\linewidth}
        \includegraphics[scale=0.28]{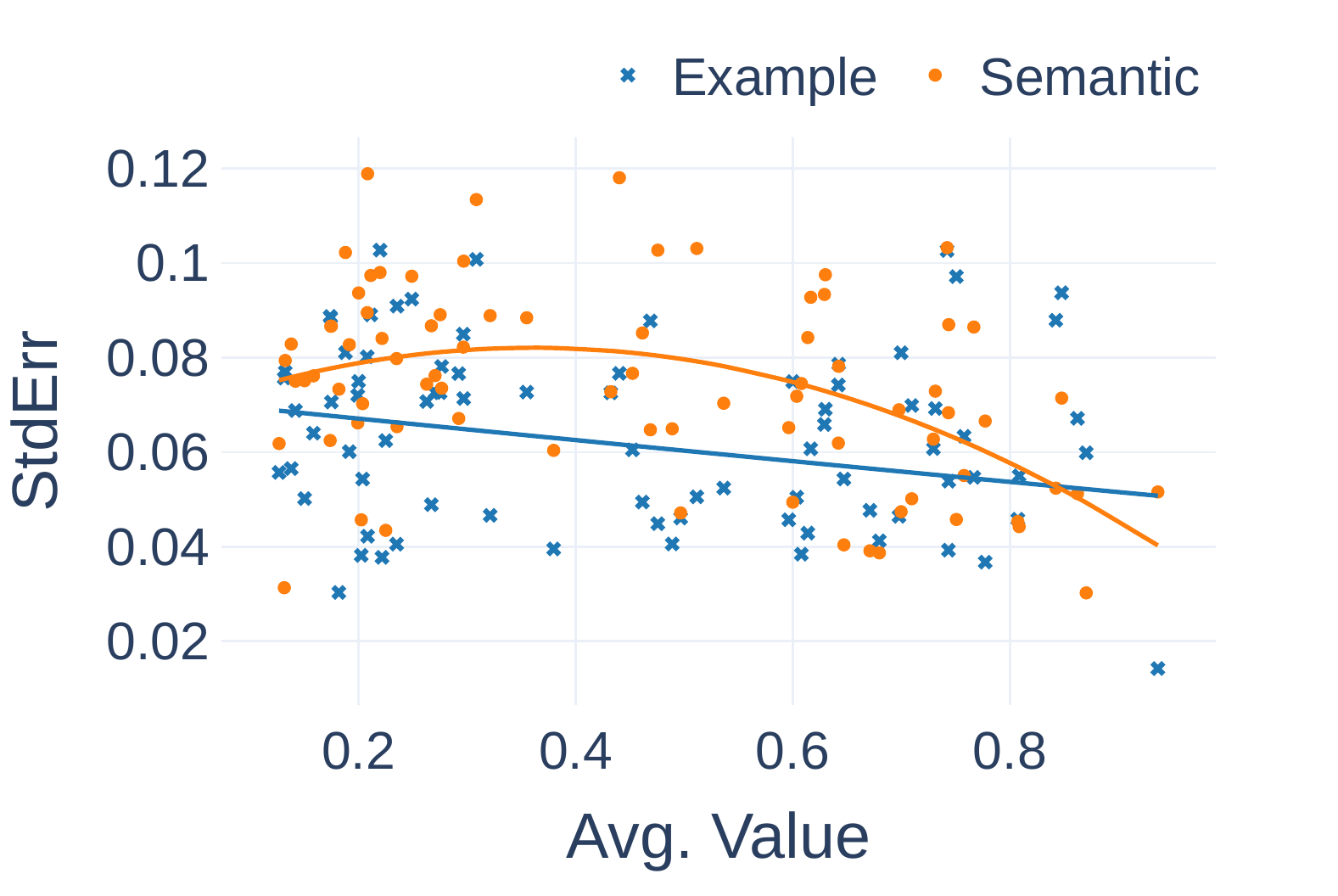}
        \caption{\textsc{toxicity} domain.}
        \label{fig:scatter-between:toxicity}
    \end{subfigure}
    \begin{subfigure}{.32\linewidth}
        \includegraphics[scale=0.28]{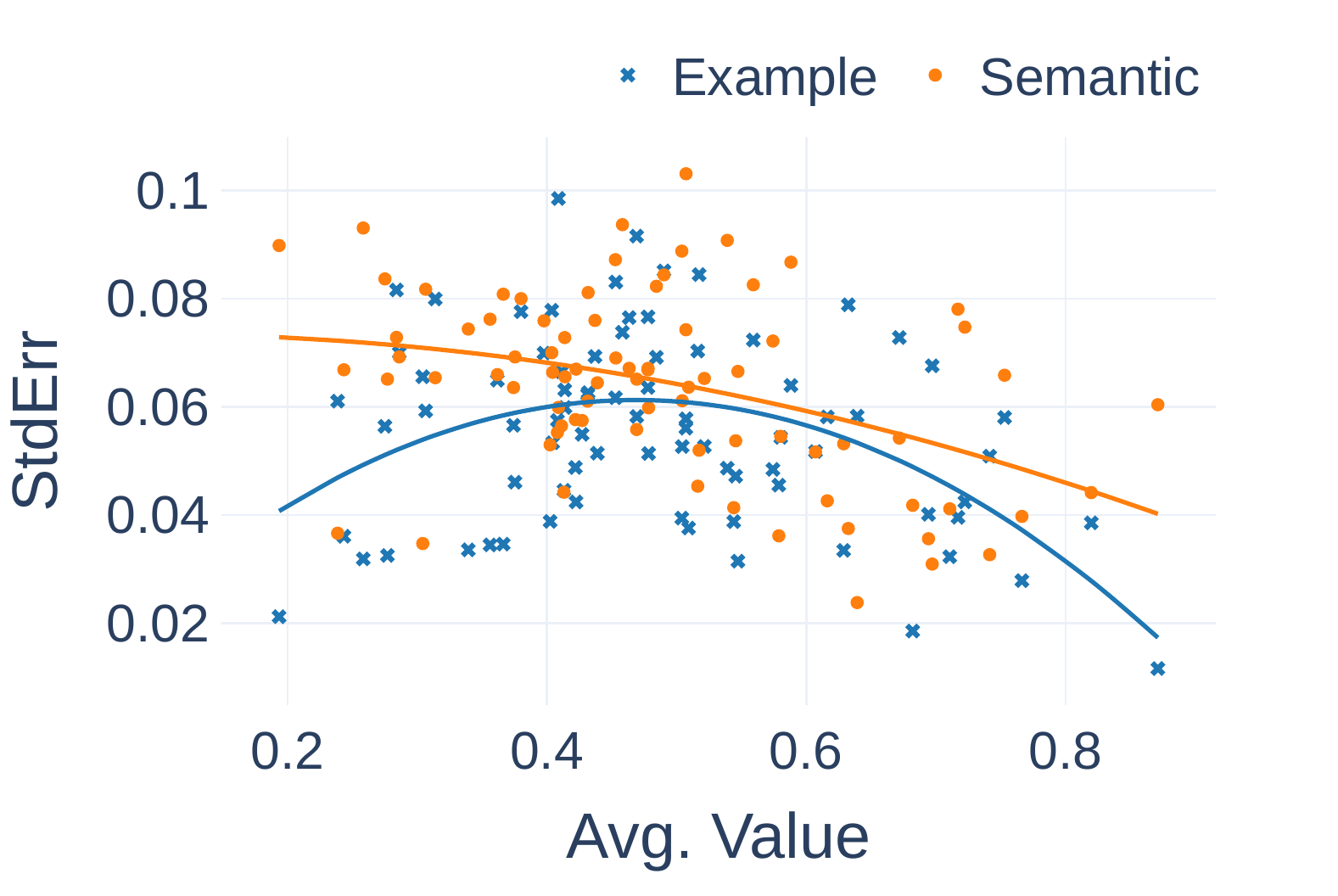}
        \caption{\textsc{satiety} domain.}
        \label{fig:scatter-between:satiety}
    \end{subfigure}
    \begin{subfigure}{.32\linewidth}
        \includegraphics[scale=0.28]{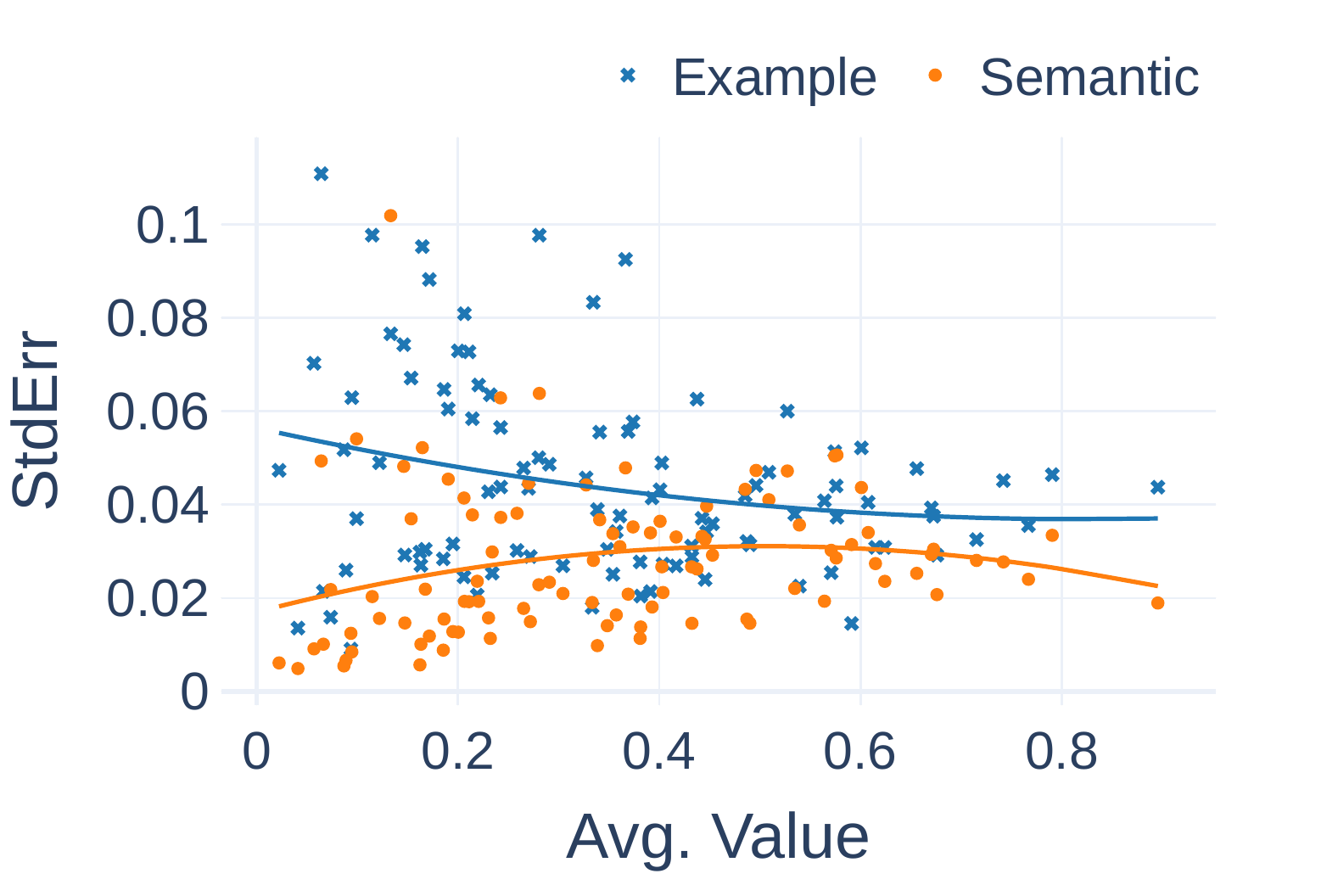}
        \caption{\textsc{age} domain.}
        \label{fig:scatter-between:age}
    \end{subfigure}
    \caption{Scatter plots of disagreement between workers (as measured by standard error) for each item plotted against the mean annotated value of each item. Trendlines represent a fit with a degree 2 polynomial.}
    \label{fig:scatter-between}
\end{figure}

We observed a statistically significant decrease in value disagreement across annotators for the \textsc{toxicity} and \textsc{satiety} domains, providing support for hypothesis H1-a. However, we observed a statistically significant increase in disagreement across annotators for the \textsc{age} domain, which contradicts H1-a. We then plotted the disagreement (standard error) in both conditions for each item against the mean value across both conditions in each domain to understand the behavioral differences we see with the age domain as shown in Figure~\ref{fig:scatter-between}. 

We find that the pattern for disagreement in the \textsc{semantic} condition is consistent with behavior observed in prior work~\cite{welty2019metrology} for similar domains with subjectivity and uncertainty. However, we note that overall disagreement between annotators was lower in the \textsc{age} domain compared to the other two domains. \chm{We also noted that scale utilization was similar in both conditions for the \textsc{toxicity} and \textsc{satiety} domains, exhibiting a slightly increase in utilization of the full scale in the \textsc{example} condition. Prior work in psychology has shown that increased spacing of items has relatively minimal effect on accurate placement when items are discriminable~\cite{Stewart2005AbsoluteIB} so we don't expect this slight increase in scale utilization to affect disagreement levels. However, opposite to the other domains, the utilization of the scale in the \textsc{age} domain was 10\% \textit{lower} for the \textsc{example} condition. We hypothesize that unlike the \textsc{toxicity} and \textsc{satiety} domains, estimating age from appearance is a domain where a numeric age scale is actually more consistently understood by human annotators,} thus example anchors provide no further benefit to annotators in understanding the scale. \chm{The scatter plots in Figure~\ref{fig:scatter-between:age} indicate that uncertainty for younger subjects was much higher in the \textsc{example} condition. Combined with the lower scale utilization we observed for \textsc{example}, we hypothesize that uncertainty about judging exact age is higher for older subjects.} As we only show example-based anchors in the \textsc{example} condition, this \chm{increased uncertainty about the reference images depicting older subjects may have resulted in more hesitation to use the higher values on the scale.} This suggests that: (1) comparisons with anchor examples mostly benefit cases where shared understanding of the scale is low, and (2) example-based anchoring should be used in \textit{addition} to semantic anchors as \textit{only} using example anchors can be detrimental to consistency if the domain is one where the semantic scale has a high degree of shared understanding already. Drawing from this experiment, our full Goldilocks annotation process uses both example-based anchors and semantic anchors to frame the scale. 

\subsection{Study 2: Evaluating Consistency Over Time Within Annotator} \label{sec:consistency-within}

For our second experiment, we explored the effect on self-consistency resulting from including an annotator's own past annotations as additional reference examples augmenting an initial seed set (H1-b). \chm{The example-based \textbf{SV-EA} interface was used for this experiment, with each annotator was assigned one of the two conditions:} \textsc{control}, where only the seed set examples was used \chm{for} reference anchors; or \textsc{augment}, where an annotator's own \chm{past annotations in the same session were included along the seed examples as references. Since we are interested mainly in the effect on self-consistency, we reduced the initial set of seed examples to just 3 examples for each domain drawn as} a subset of the 7 example instances used in the \textsc{example} condition of the previous experiment. \chm{We took the items corresponding to the lowest, highest, and median ratings.}

The items in each domain were shuffled and then partitioned into sequences of size 20, resulting in 5 sequences for the \textsc{toxicity} and \textsc{age} domains and 4 sequences for the \textsc{satiety} domain. Each annotator was given interface training and then subsequently tasked with annotating one of the sequences (of 20 items). To probe for changes in the rating of an item, we replaced the 10th and 20th items in each sequence above with repeats of the first item, which we will refer to as the probe item. When the probe item is annotated in the \textsc{augment} condition, the annotator's own past annotation for the probe item will be withheld from the set of reference items. We measure $\Delta_1$ as the size of the value change between the first and second annotation attempts of the probe item and $\Delta_2$ as the size of the value change between the second and third annotation attempts of the probe item.

\subsubsection{Results}

\begin{figure}[tbh]
    \centering
    \begin{subfigure}{.32\linewidth}
        \includegraphics[scale=0.28]{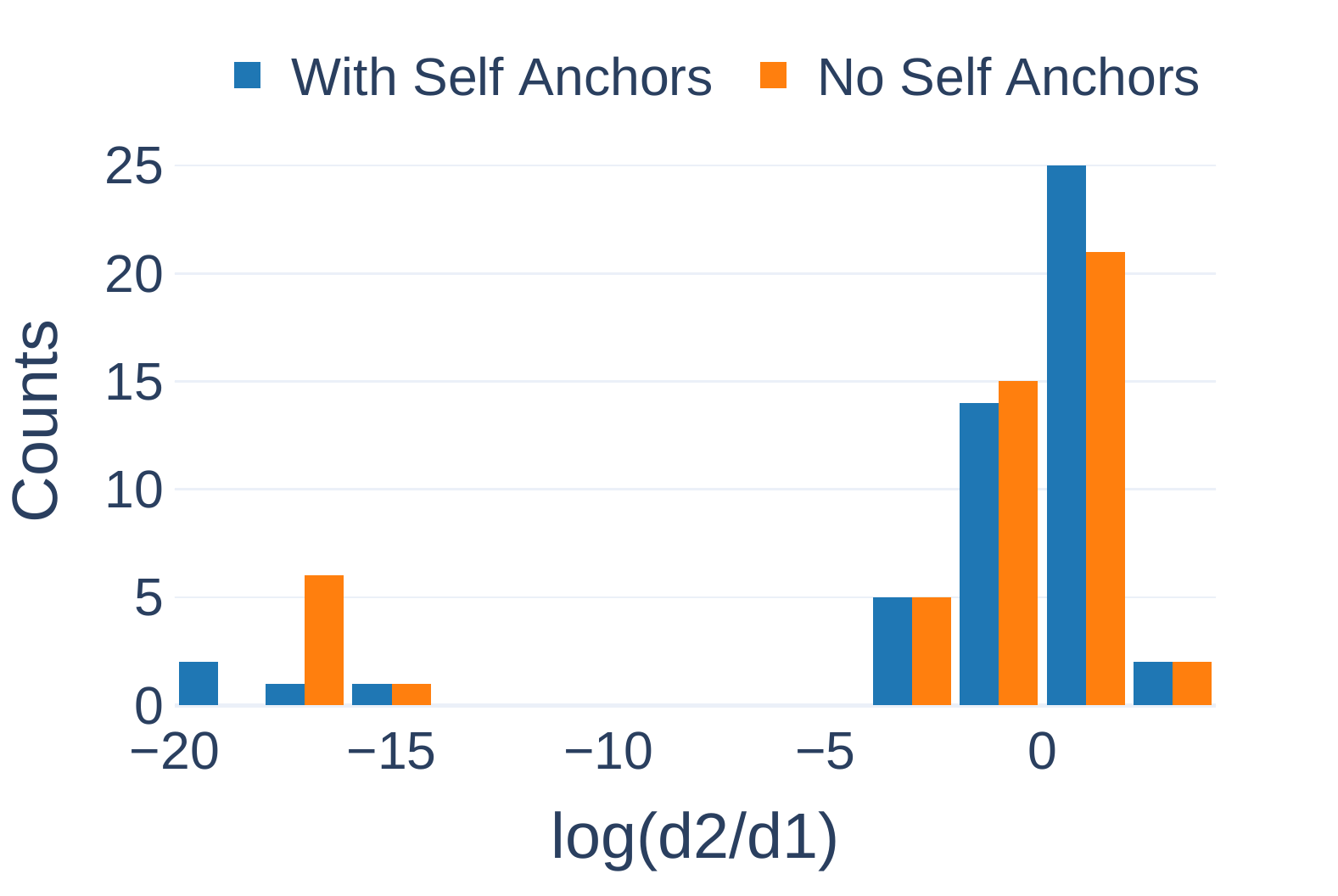}
        \caption{\textsc{toxicity} domain.}
        \label{fig:bar-within:toxicity}
    \end{subfigure}
    \begin{subfigure}{.32\linewidth}
        \includegraphics[scale=0.28]{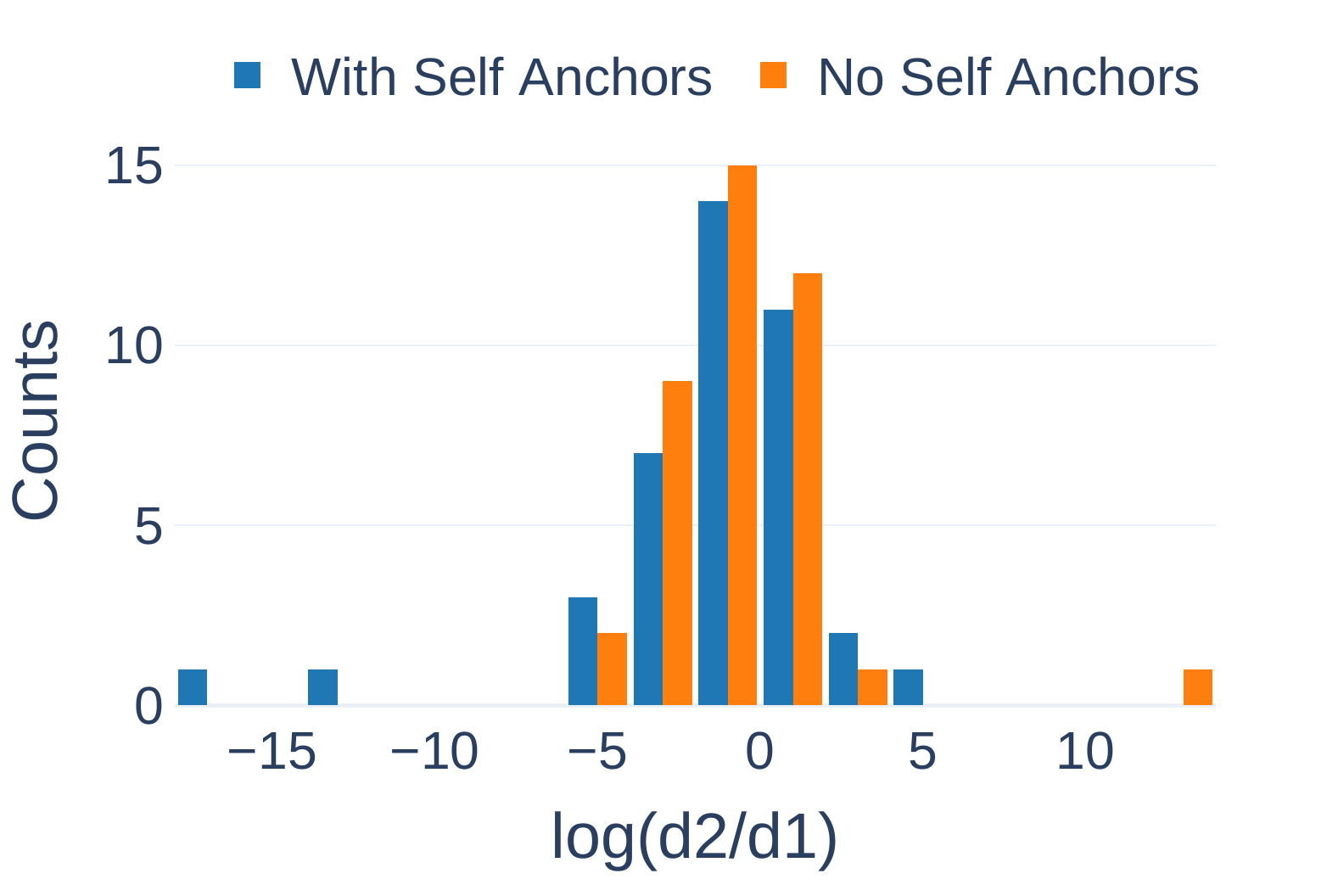}
        \caption{\textsc{satiety} domain.}
        \label{fig:bar-within:satiety}
    \end{subfigure}
    \begin{subfigure}{.32\linewidth}
        \includegraphics[scale=0.28]{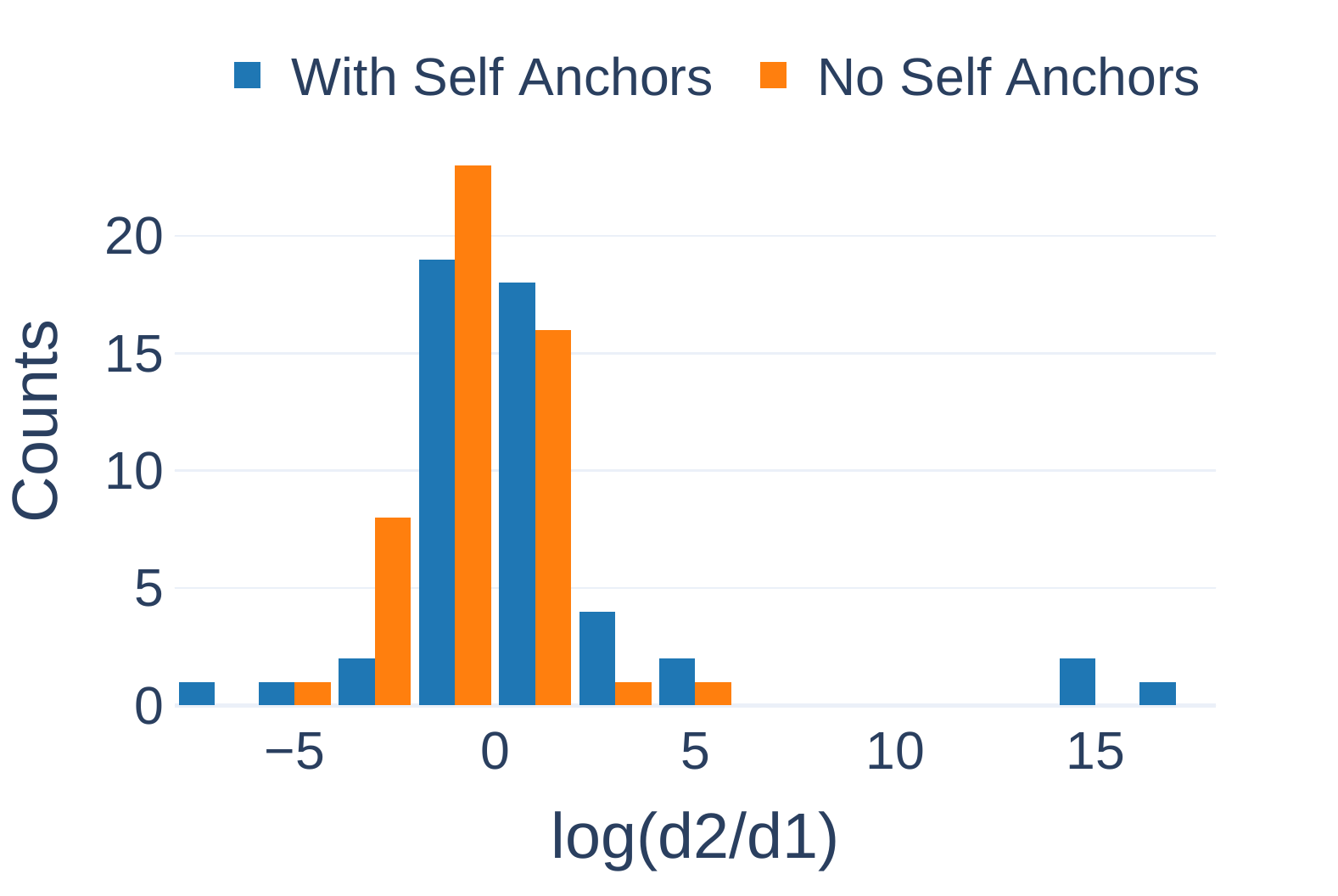}
        \caption{\textsc{age} domain.}
        \label{fig:bar-within:age}
    \end{subfigure}
    \caption{Histogram of distance ratios between first re-annotation and second re-annotation of the probe item on a log scale. Negative values indicate more decrease in disagreement with the annotator's own answers while positive values indicate more increase in disagreement with the annotator's own answers. Ratios were smoothed using Laplace smoothing with $\epsilon = 10^{-8}$.}
    \label{fig:bar-within}
\end{figure}

\chm{From Table~\ref{tab:breakdowns_delta} we can see that for most domain condition pairs, the absolute amount of an annotator's disageement with their past rating tends to exhibit a natural decrease as they get familiarized with the scale. Since the magnitude of initial self-disagreement for the probe item varies for each annotator, comparing absolute change in self-disagreement can be misleading as the same \textit{proportional} change in self-disagreement will reflect as a larger \textit{absolute} change. To account for these factors, we instead look to the self-disagreement \textit{ratio} ($\Delta_2 / \Delta_1$) as a measurement for the proportional decrease (or increase) in self-disagreement. Ratios below 1 indicate that self-disagreement has decreased while those above 1 indicate an increase. In Figure~\ref{fig:bar-within}, we show a histogram of this ratio on a log-scale for each condition in this study.} 

Our first step is understanding whether self consistency improves over time simply from doing the task and being exposed to more examples. We conducted a sign test for each of the task domains and find that in the \textsc{toxicity} domain, self consistency does improve over time (P < 0.005) for both \textsc{control} and \textsc{augment} conditions. Self consistency was not found to have a significant across-the-board improvement in any of the other domains. Comparing across the two conditions, we did not measure significant effect on self-disagreement ratio in any of the 3 domains.

\begin{table}[t]
    \centering
    \caption{\chm{Table breakdown of the change in rating for the probe item (compared to its last most recent rating) when re-annotated for the first time ($\Delta_1$) and when re-annotated the second time ($\Delta_2$). The ``Top Avg. $\Delta$'' columns represent the averages when only considering the instances where $\Delta_1$ was among the top 30\% most uncertain.}}
    \label{tab:breakdowns_delta}
    \begin{tabular}{lccccc}
    \toprule
    \chm{\textbf{Domain}}  & \chm{\textbf{Condition}} & \chm{\textbf{Avg. $\Delta_1$}} & \chm{\textbf{Avg. $\Delta_2$}} & \chm{\textbf{Top Avg. $\Delta_1$}} & \chm{\textbf{Top Avg. $\Delta_2$}}\\ 
    \midrule
    \chm{\textsc{toxicity}} & \chm{No Self (\textsc{control})}      & \chm{0.105} & \chm{0.062}  & \chm{0.244} & \chm{0.077} \\
                            & \chm{With Self (\textsc{augment})}    & \chm{0.133} & \chm{0.090}  & \chm{0.347} & \chm{0.095}\\
    \midrule
    \chm{\textsc{satiety}}  & \chm{No Self (\textsc{control})}      & \chm{0.140} & \chm{0.086}  & \chm{0.308} & \chm{0.171}\\
                            & \chm{With Self (\textsc{augment})}    & \chm{0.126} & \chm{0.052}  & \chm{0.286} & \chm{0.027}\\
    \midrule
    \chm{\textsc{age}}      & \chm{No Self (\textsc{control})}      & \chm{0.110} & \chm{0.063}  & \chm{0.280} & \chm{0.133}\\
                            & \chm{With Self (\textsc{augment})}    & \chm{0.063} & \chm{0.066}  & \chm{0.157} & \chm{0.067}\\
    \bottomrule
    \end{tabular}
\end{table}

We then hypothesized that effect on self-consistency may not be uniform across all probe items---if an annotator already has low self-disagreement in the first re-annotation round ($\Delta_1$), it likely implies there is little uncertainty about the placement of the item and thus we shouldn't expect further improvements. \chm{Considering this, we now look at only the top 30\% `most uncertain' annotation sessions for each domain and condition combination, as sorted by decreasing $\Delta_1$. This set consists of 15 sessions for the \textsc{toxicity} and \textsc{age} domains and 12 for the \textsc{satiety} domain. In this high-disagreement subset of sessions,} we find that augmenting reference examples (\textsc{augment}) \chm{with past annotations in the session does result in a larger proportional reduction in self-disagreement (reflected through self-disagreement ratios) when compared to} \textsc{control} for both the \textsc{toxicity} and \textsc{satiety} domains. \chm{For the \textsc{satiety} domain, median proportional decrease in self-disagreement was $0.076$ ($92\%$ reduction in self disagreement) for the \textsc{augment} condition compared to $0.263$ ($74\%$ reduction) for the \textsc{control}. The median ratios were $0.190$ ($81\%$ reduction) and $0.310$ ($69\%$ reduction) respectively for the \textsc{toxicity} domain. However, the limited amount of data points in these groups means we do not have statistical power to claim significance.} Overall, we don't find sufficient support for H1-b, but we note a pattern of improvement in self-consistency for items with high initial self-disagreement when including an annotator's own prior annotations as additional references. Similar to the previous section, we were unable to observe benefit of augmenting reference examples on the \textsc{age} domain, likely due to the already limited utility of reference examples in this domain.

\subsection{Study 3: Evaluating Range Annotation} \label{sec:ranges}
For the final experiment, we explored how robustly ranges produced by the two-step annotation process in Goldilocks reflect properties of relationships between items. \chm{In this experiment, annotators were asked to annotate a sequence of items using the full Goldilocks two-step annotation process (using the \textbf{R-HA} design shown in Figure~\ref{fig:scale}).} The annotation experiments were conducted on the \textsc{toxicity} and \textsc{satiety} domains with sequences generated by shuffling each dataset and partitioning the dataset into groups of size 10, resulting in 10 and 8 groups respectively for the two domains. We then recruited 5 annotators to annotate each sequence in each of the domains. 

At the start of the task, each annotator was first trained on how to use the two-step annotation system described earlier in Section~\ref{sec:twostep} by annotating a sample task with guidance given during each step.  
After the annotator completes the training example item, they then proceed to annotate the assigned sequence of 10 task items. To seed the initial reference examples, we used the same reference anchors as used in the first experiment. We also included each annotator's own annotations as anchors during their annotation in a similar way as the \textsc{augment} condition in the second experiment.

\subsubsection{\chm{Establishing Pairwise Relationship Distributions}} \label{sec:pairwise-relationship-distributions}

\chm{In order to measure ground truth distributions over the pairwise relationships, we recruited separate workers and used the \textbf{Pairwise} design to directly collect pairwise judgments on relationships ($>$,$<$,$\approx$) between all pairs of items in each group. Distributions across the 3 relationship types were then created by counting the proportion of annotators indicating each type of relationship across for each pair of items. These distributions reflect the degree of disagreement among annotators for the pairwise relationship.} 

\chm{We then considered how one would recover similar distributions across relationships for pairs of items using the traditional approach of single-value absolute rating scales based on semantic anchors, creating two alternative baselines. Since the traditional approach cannot simultaneously elicit item ambiguity and agreement, producing a similar distribution would involve a tradeoff.}

\chm{For the \textbf{Direct} baseline, we assume that there is no item-level ambiguity, meaning that even local pairwise comparisons can be made by directly comparing the raw values from the absolute rating. For example, we count an annotator as indicating a ``$>$'' relationship on a pair $(a, b)$ if their single rating scores indicate $r_a > r_b$. One can generally expect this to be reliable when $a$ and $b$ are far apart on the scale but it can be much less reliable for close neighbors. }

\chm{For the \textbf{Infer} baseline, we assume that all disagreement observed between annotators reflects the ambiguity of the item. In this case, we aggregate the individual ratings into a \textit{single} 95\% confidence interval for each item by measuring the mean and standard error between these samples. We then infer the relationship between of a pair of items by comparing the confidence intervals, treating overlapping intervals as indicating a relationship of `indistinguishable ($\approx$)'. In this case, the distribution across relationships for a pair would see all the probability mass allocated to the single relationship measurement produced by the comparison. }

\chm{Finally, with Goldilocks annotation, we have range evaluations on a per-annotator granularity. For each annotator, we can use their range labels to find the relationship between two items, treating overlaps as indicating $\approx$. We can then produce a distribution by counting the proportion of annotators indicating each relationship. With Goldilocks we don't need to make tradeoffs between measuring item ambiguity and agreement.}

\subsubsection{Results: Recovering relationships between items}

\begin{table}[t]
    \centering
    \caption{Comparing the quality of the pairwise relationship distributions as recovered by (1) \textbf{range}s collected in Goldilocks, \chm{(2) \textbf{direct}ly comparing raw values picked by each annotator, and (3) indirectly using ranges \textbf{infer}red from the 95\% confidence intervals. Details in~\ref{sec:pairwise-relationship-distributions}.} Wasserstein distance to the ground truth distribution (collected \chm{directly using pairwise comparisons}) was computed for each case. Goldilocks ranges produce distributions the closest (least distance) to the ground truth. }
    \label{tab:pair_relationship_distance}
    \begin{tabular}{lccc}
    \toprule
    \textbf{Domain}  & \textbf{Avg. WD (Range)} & \textbf{Avg. WD (Direct)} & \textbf{Avg. WD (Infer)} \\ 
    \midrule
    \textsc{toxicity} & \textbf{0.332314}  & 0.366944 & 0.450556\\
    \midrule
    \textsc{satiety}  & \textbf{0.424352}  & 0.449444 & 0.597222\\
    \bottomrule
    \end{tabular}
\end{table}

To compare and quantify how robustly each of these methods recovers relationships, we \chm{measured the Wasserstein distance between relationship distributions for each of the 3 approaches in \ref{sec:pairwise-relationship-distributions} and the ground truth relationship distributions collected through pairwise comparative rating.} Table~\ref{tab:pair_relationship_distance} shows that among the 3 methods to produce distributions over pairwise relationships, \chm{recovering distributions using range labels most accurately agrees with the ground truth distribution}, supporting H2-a. 

\chm{We found that using inter-annotator agreement to infer the inherent ambiguity (referred to by prior works as ``resolution'') of items results in an over-estimate of the amount of ambiguity.} In the \textsc{toxicity} domain, 43.5\% of the relationships that \chm{were} distinguished in the ground truth \chm{distribution collected directly through pairwise comparisons} were inferred to be ``indistinguishable'', with this ratio as high as 68.1\% in the \textsc{satiety} domain. In contrast, ranges \chm{over-estimate ambiguity (under-estimating resolution) only} about half as often, with 22.1\% and 30.9\% respectively. This supports the idea that ranges are a better model of resolution (H2-b).

\subsubsection{Results: Comparing aggregation uncertainty with range sizes}

\begin{figure}[tb]
    \centering
    \begin{subfigure}{.49\linewidth}
        \includegraphics[scale=0.40]{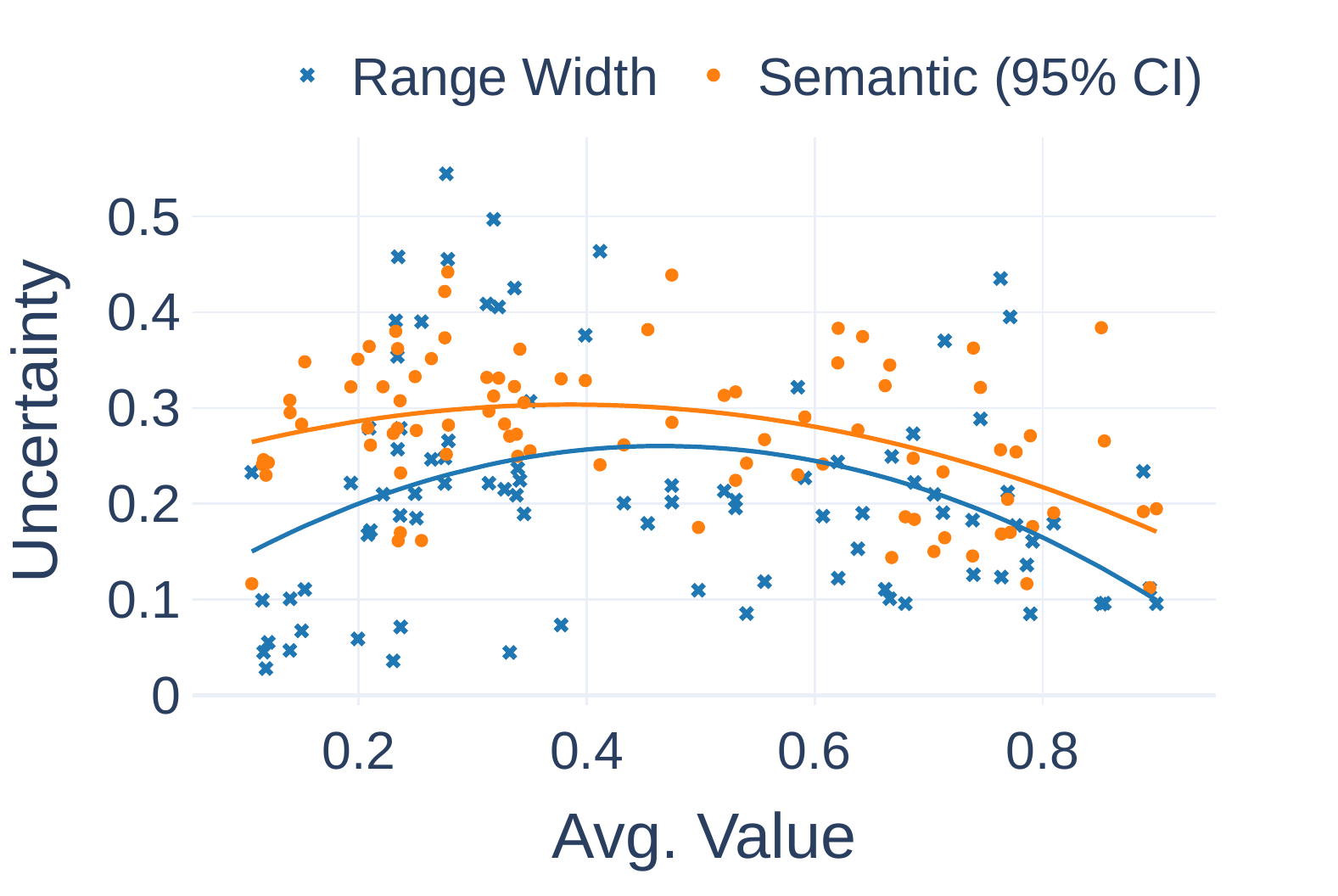}
        \caption{\textsc{toxicity} domain.}
        \label{fig:sizes:toxicity}
    \end{subfigure}
    \begin{subfigure}{.49\linewidth}
        \includegraphics[scale=0.40]{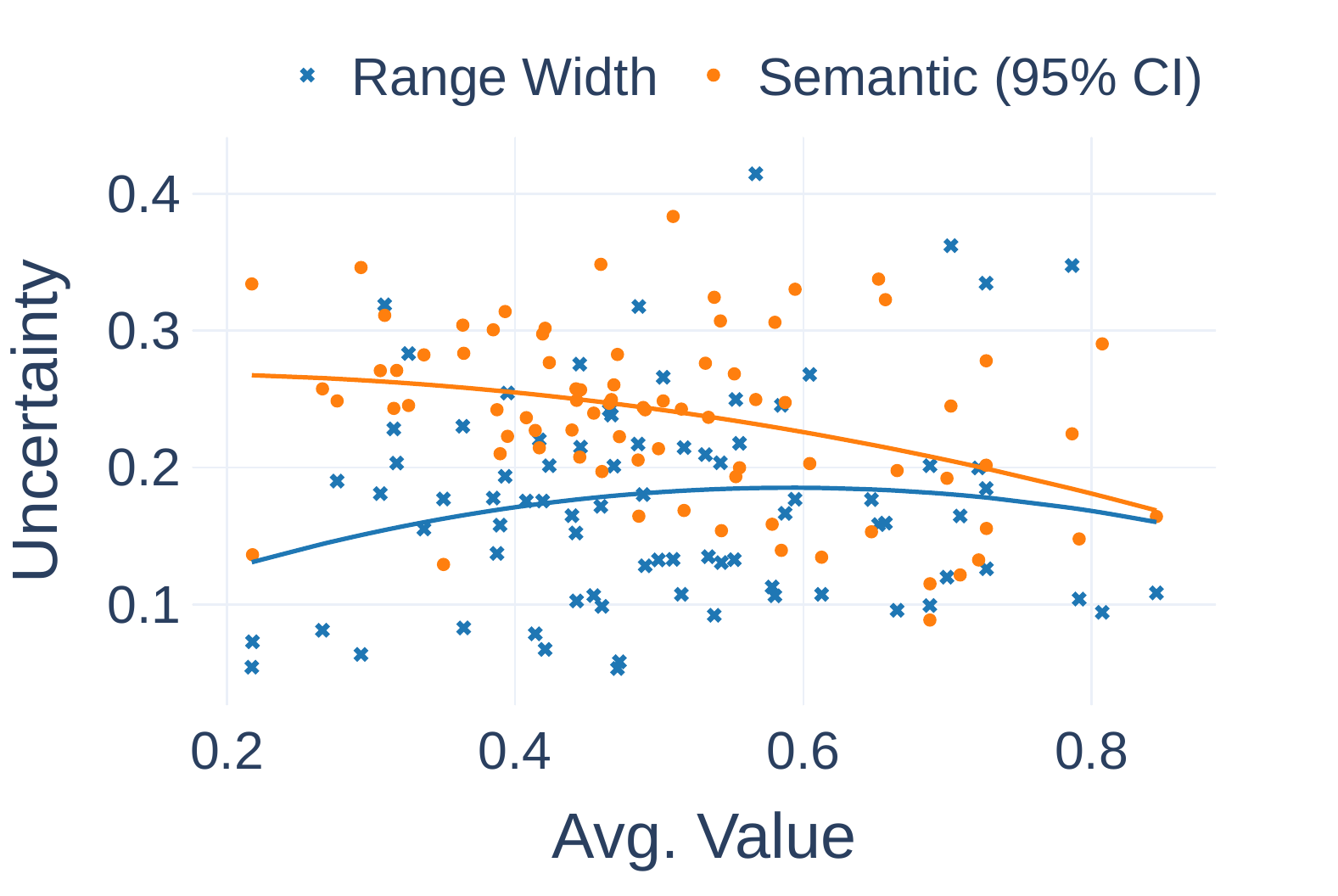}
        \caption{\textsc{satiety} domain.}
        \label{fig:sizes:satiety}
    \end{subfigure}
    \caption{Comparison of uncertainty measured as range sizes from Goldilocks annotation with uncertainty measured through standard error confidence intervals from traditional single-value semantic scale annotation. Trendlines represent a fit with degree 2 polynomials.}
    \label{fig:sizes}
\end{figure}

Finally, we explored differences in the type of uncertainty measured through Goldilocks annotation ranges sizes with uncertainty measured by confidence intervals of annotations using semantic scales. We hypothesize that since ranges focus on capturing resolution (distinguishability against peers) of items, the resulting uncertainty represented by the size of ranges will be different than uncertainty represented by inter-annotator disagreement metrics, though the two may still be related.

First we look at the behavior of the two kinds of uncertainty measurements across the range of values on the scale. Figure~\ref{fig:sizes} plots the two kinds of uncertainty: average size of ranges and 95\% confidence intervals for semantic scale annotation values. We find that overall range sizes represent uncertainty lower than that measured by 95\% confidence intervals from aggregating semantic scale annotation (P < 0.001). This makes intuitive sense as we would expect item level resolution to be a tighter uncertainty. We also find that in the \textsc{toxicity} domain, both types of uncertainty behave similarly with respect to extreme values on the scale corresponding to lower values of uncertainty in definitions. In the \textsc{satiety} domain, however, we found that lower values (corresponding to foods depictions that are less satiating) corresponded to larger uncertainty in the form of disagreement but not with range sizes. We think this may result from higher disagreement about what foods are not satiating among different annotators but with annotators each confident about their own determination of satiety (high resolution/distinguishablity of items).

Looking at correlation between the values produced by the two types of uncertainty, we observe only very weak correlation between range sizes and confidence intervals (scaled standard error) for both \textsc{toxicity} and \textsc{satiety} domains with $R^2 < 0.01$ in both domains. This indicates that the uncertainty we measure with ranges does not have significant correlation with inter-annotator disagreement measures like standard error (RQ3). We note that with range annotations, inter-annotator disagreement measures can be further computed for the range bounds themselves to evaluate disagreement separately from item uncertainty (resolution) captured by ranges. However, as single-value semantic scalar annotations can't facilitate separation of the two uncertainty types, we are unable to make direct comparisons. 

\section{Discussion}

In the prior sections, we demonstrate that the ideas of grounding absolute rating scales with examples and explicitly capturing item-level measurement resolution can be beneficial for more consistent and robust annotation of subjective domains lacking shared understanding of absolute ratings scales. In this section, we will discuss some of the other considerations in adapting Goldilocks as a full annotation technique, including examining the annotation efficiency (in terms of work time) of Goldilocks compared to hybrid application of traditional methods and envisioning how Goldilocks may be scaled up to multiple annotation sessions using iterative-improvement processes. We will also discuss limitations of the Goldilocks process and potential avenues for future work.

\subsection{\chm{Annotation Efficiency and Cost of Range Annotations}}

\chm{One of the main advantages of the Goldilocks annotation process is the ability to capture item-level ambiguity and disagreement between annotators simultaneously through the use of range annotations. 
However, separating these sources of uncertainty comes at an extra cost for the data collection process---even though range bounds in Goldilocks can be collected with low overhead compared to traditional absolute rating, the tasks can be more work for the requester to set up. This presents a tradeoff for practitioners when deciding whether the higher quality of data is worth the cost. Prior work simulating data annotation tasks inspired by measuring objective properties has shown that, given a fixed budget, some learning algorithms actually benefit more from a larger amount of lower-quality annotations on novel examples rather than higher-quality annotations on fewer items~\cite{Lin2014ToRO}. 
Indeed, for these tasks where disagreement is likely caused by noisy perception, it's likely that a practitioner will see relatively little benefit by separating item-level ambiguity from annotator disagreement. However, with the rising demand for training data in domains that involves subjectivity or nuance, understanding and accounting for sources of uncertainty and limitations within the data itself has become increasingly important for building models that are \textit{trustworthy} rather than just more \textit{performant}~\cite{Bhatt2020UncertaintyAA}. Separating disagreement from inherent ambiguity using range-based annotation can also offer better transparency about the annotation process and data produced, allowing for the potential to diagnose model limitations and human biases even into the future. In these cases, the higher cost of setting up Goldilocks annotations can be justified by the richer information that can be derived from range-based rating data. }

\chm{Of course, Goldilocks is not the only approach to capture both item-level ambiguity and disagreement. It is possible to use traditional absolute and comparative rating to separately collect scalar annotations and pairwise comparisons to recreate absolute rating estimates and pairwise relationship distributions. We also wanted to understand whether Goldilocks can provide efficiency benefits when compared to hypothetical hybrid approches using only a combination of traditional annotation interfaces. We look at the work time taken by crowd workers in our various experiments to extrapolate the effort necessary for such an approach.} Assuming a task group size of 10 items, we find that the Goldilocks two-step workflow results in a median work time (including both training and annotation) of $429.5$s per worker per task group on the \textsc{satiety} domain and $592.5$s per worker per task group on the \textsc{toxicity} domain. Collecting only single value rating annotations with Likert-style anchors takes a median work time of $307.5$s per worker per task group on the \textsc{satiety} domain and $238$s per worker per task group in the \textsc{toxicity} domain. Finally, comparative rating on a group of size 10 implies 45 pairwise comparisons to capture full pairwise relationships, which takes a median time of $513.5$s per worker per group and $502$s per worker per group for the two domains respectively. Thus we expect that at the same level of redundancy for annotations, Goldilocks can be 20-48\% more efficient \chm{through the use of our two-step range-based annotation that collects ratings and relationship distributions together. Consistency improvements of Goldilocks may be able to push efficiency further in practice by requiring a lower amount of redundancy to achieve the same level of agreement.}

\subsection{Goldilocks and Iterative Improvement} \label{sec:iterative-improvement}

So far in this paper we have examined the ideas presented in Goldilocks only for single annotation sessions \chm{where we didn't need to update the anchor examples beyond incorporating an annotator's own ratings. In order to scale up to larger datasets, it becomes necessary to perform annotations over multiple sessions which involve using aggregation approaches to iteratively construct an updated set of anchors.} To achieve this we envision a process based on the idea of iterative improvement~\cite{Little2010iterativeimprovement}.

In each round of iteration, a group of annotators individually annotate a subset of the dataset, sharing a `seed' set of \chm{anchor} examples used to ground the interpretation of the scale, with their own annotations \chm{also incorporated} as they progress along the annotation session. Once all annotators have completed the session, the annotations collected will be aggregated into a new set of seed examples used to ground the next round of iteration. In addition to progressively annotating new examples, this iterative process may also be used to revise past annotations, such as those created during the cold start process. \chm{This can be accomplished} by \chm{first} removing the items to be revised from the set of grounding examples and \chm{then} re-annotating them \chm{as new items} in an iteration. This process of periodically aggregating annotations and then re-seeding \chm{anchor} examples \chm{can serve as a method to scale up annotations while ensuring a stable scale as annotators place items.} 

We believe that this represents a feasible design for scaling up annotation, and we envision further work can be done to explore options for aggregation and re-annotation strategies as well as evaluate their effectiveness. We also see potential for using iterative improvement as way to dynamically re-calibrate the definition of scales to account for distributional shifts over time. For example, a scale that can dynamically adapt to improving quality of machine summarization systems can be adapted as a living benchmark. We think the ideas presented in Goldilocks for single annotation sessions provide a first step into building an effective iterative workflow.

\subsection{Limitations and Future Work}
While Goldilocks \chm{provides a path to more consistent scalar annotation that also captures uncertainty, we also recognize that the current design is still subject to some limitations which we believe can be good avenues for future work.}

\subsubsection{\chm{Creating High Quality Seeds in Cold Start}} \label{sec:discussion-cold-start}

\chm{The cold start process in Goldilocks provides a way to generate the initial seed set of grounding examples that enable the comparisons and consistency benefits of Goldilocks. However, the quality of this initial set of seed examples can also influence whether consistency benefits can be realized. We observed some of these limitations when experimenting with example-based anchors in the \textsc{age} domain. A good seed set should consist of examples that achieve both good coverage of the scale and have low ambiguity themselves. When the seed set achieves good coverage over the scale, the comparative process can allow seed examples that are distinguishable to quickly be excluded from the range of the annotated item, resulting in measurement resolution that mainly depends on the number of examples in the seed set. However, a set of examples that is not representative of the full range of items to be ranked can lead to issues of scale drift when these examples (that annotators may desire to rate higher or lower than the current implied bounds of the scale) are encountered in the future. The current cold start process provides some mitigation to the issue of representativeness by incorporating a `resampling and replace' phase to increase the diversity of items in the seed set. However, for sufficiently large datasets this may not be enough to capture rare items that are also outliers for the scale. For future work, we envision enabling the ability for annotators to rescale the visible scale itself through an interaction similar to zooming in or out, allowing the annotation of items that lie outside the current extremes of the scale when they are encountered. }

\chm{Another current limitation of the cold start process is that the cold start design cannot effectively capture item ambiguity as we only elicit a single label for each reference item. In pilot studies we found it infeasible to introduce ranges into the cold start process as there are no anchors to compare against to effectively determine these ranges. It is possible to have suboptimal seed sets where the seed items can have high ambiguity themselves, thus acting as a lower bound on range sizes. We hypothesize that the iterative improvement process in ~\ref{sec:iterative-improvement} may offer a way to limit the impact of the cold start seeds if we can conduct subsequent annotation rounds where we can instead seed with regular annotated range data, though we leave exploration of this to a future study.}

\subsubsection{Addressing Long-form \chm{Tasks} and Context} 

Some common tasks where crowd scalar ratings are desirable, such as evaluating \chm{relevance, conciseness, fluency, or faithfulness} of summaries produced by text summarization models, can depend on understanding long-form context (e.g., a news article) or even multiple documents~\cite{Fabbri2020SummEvalRS}. While \chm{we have shown that} Goldilocks can support \chm{annotation domains based on} small amounts of text (1-2 sentences) using a similar interface as the one used for images, \chm{long-form} text will require a different design for conducting comparisons both with the global scale and local neighborhood.

Additionally, interactions in Goldilocks assume that items can be compared against other items in the same dataset. However, when rating items with context, such as summarization or translation, it is likely that reasonable comparisons can only be made with certain other items sharing the same context (i.e., alternate summaries/translations of the same source). A potential avenue for future work extending Goldilocks may exist in introducing virtual views to the Goldilocks scale that enable contextual comparisons on the scale by only exposing items sharing the same context. Future work on an algorithm for determining optimal global example anchors could also take into account aspects that could make comparison easier, such as similarity to the item being annotated.

\subsubsection{Working with Density}

One of the strengths of Goldilocks is the ability to use past annotations from any source, including data from existing datasets to establish grounding for a scale. By providing past annotations from a dataset as reference examples, it will be possible to augment the dataset in a way that is consistent with past examples but also doesn't require building complex rubrics. However, as the set of past annotations increases, it poses potential problems for the local comparison aspect of the Goldilocks annotation process. There are practical limitations on how fine adjustments can be on a slider-based scale, so as regions on the scale become densely populated by examples, it becomes harder to use local comparisons to find precise upper and lower bounds in those regions. Even small adjustments in a dense region can mean moving across many reference points.

One potential solution to the density problem could come from allowing the scale to be itself scaled\chm{, similar to that proposed in~\ref{sec:discussion-cold-start}. Initially the full view of the scale is presented along with global anchors for coarse navigation. As an annotator narrows down on a dense region,} they can increase the zoom level of the annotation scale to span \chm{just} the dense region across the entire width of the scale, increasing the amount of space and in turn reducing interaction issues caused by density. \chm{New global anchors can be selected to allow for quick navigation at the new zoom level.}

\section{Conclusion}
In this paper, we present and evaluate Goldilocks, a novel technique to elicit scalar annotations using the crowd that improves on consistency and captures pairwise relationships more robustly. We show that by prior examples can be used as anchors to ground otherwise abstract absolute rating scales (such as semantic or Likert scales) leading to more consistent interpretation between workers. We find that including an annotator's past annotations in a session can lead to more self consistency on items that have high initial uncertainty. Finally, we show that introducing range annotation into absolute rating can enable simultaneous elicitation of both perceived ambiguity on a per-annotator scale while also capturing inter-annotator disagreement. This simultaneous measurement enables a better recovery of pairwise relationship distributions.


\bibliographystyle{ACM-Reference-Format}
\bibliography{biblio}


\end{document}